\documentclass[natbib,smallextended]{svjour3}
\smartqed  

\usepackage{graphicx}
\usepackage{amsmath}
\usepackage{aas_macros}
\usepackage[colorlinks=true,urlcolor=blue,citecolor=blue,linkcolor=blue,breaklinks]{hyperref}

\usepackage[colorlinks=true,urlcolor=blue,citecolor=blue,linkcolor=blue]{hyperref}

\makeatletter
\newcommand{\bibnote}[2]{\@namedef{#1note}{#2}}
\newcommand{\biblink}[2]{\@namedef{#1link}{#2}}
\makeatother

\newcommand{\edt}[1]{{\bf{#1}}}

\newcommand{\kms}{km\,s$^{-1}$}

\newcommand{\dd}{\mathrm{d}}

\newcommand{\dnu}{\ensuremath{\dd \nu}}
\newcommand{\Inu}{\ensuremath{I_\nu}}
\newcommand{\jnu}{\ensuremath{j_\nu}}
\newcommand{\Snu}{\ensuremath{S_\nu}}
\newcommand{\Jnu}{\ensuremath{J_\nu}}
\newcommand{\Bnu}{\ensuremath{B_\nu}}
\newcommand{\anu}{\ensuremath{\alpha_\nu}}
\newcommand{\tnu}{\ensuremath{\tau_\nu}}

\newcommand{\grad}{\ensuremath{\vec{\nabla}}}

\newcommand{\vn}{\ensuremath{\vec{n}}}
\newcommand{\vp}{\ensuremath{\vec{p}}}
\newcommand{\vv}{\ensuremath{\vec{v}}}
\newcommand{\vg}{\ensuremath{\vec{g}}}

\newcommand{\vF}{\ensuremath{\vec{F}}}
\newcommand{\vJ}{\ensuremath{\vec{J}}}
\newcommand{\vP}{\ensuremath{\vec{P}}}
\newcommand{\vB}{\ensuremath{\vec{B}}}

\newcommand{\Fnu}{\vF_\nu}
\newcommand{\knu}{\kappa_\nu}
\newcommand{\kR}{\ensuremath{\kappa_\mathrm{R}}}
\newcommand{\tR}{\ensuremath{\tau_\mathrm{R}}}
\newcommand{\kB}{\ensuremath{\kappa_B}}

\newcommand{\rad}{\ensuremath\mathrm{rad}}
\newcommand{\Qrad}{\ensuremath{Q_\rad}}

\newcommand{\divF}{\ensuremath{\nabla \cdot \vF}}

\newcommand{\be}{\begin{equation}}
\newcommand{\ee}{\end{equation}}
\newcommand{\bea}{\begin{eqnarray}}
\newcommand{\eea}{\end{eqnarray}}

\DeclareRobustCommand{\ion}[2]{\textup{#1\,\textsc{\lowercase{#2}}}}
\newcommand{\HI}{\ion{H}{i}}
\newcommand{\HII}{\ion{H}{ii}}
\newcommand{\CaII}{\ion{Ca}{ii}}
\newcommand{\MgII}{\ion{Mg}{ii}}
\newcommand{\HeI}{\ion{He}{i}}
\newcommand{\HeII}{\ion{He}{ii}}

\newcommand{\rme}{\ensuremath{\mathrm{e}}}
\newcommand{\rml}{\ensuremath{\mathrm{l}}}
\newcommand{\nH}{\ensuremath{n_\mathrm{H}}}
\newcommand{\nel}{\ensuremath{n_\rme}}

\newcommand\nijk{\ensuremath{n_{ijk}}}
\newcommand\kb{\ensuremath{k_\mathrm{B} }}
\newcommand\nne{\ensuremath{n_\mathrm{e}}}

\journalname{Living Reviews in Solar Physics}

\begin{document}

\title{Radiation hydrodynamics in simulations of the solar atmosphere}

\author{Jorrit Leenaarts}

\institute{J. Leenaarts\at
Institute for Solar Physics, Department of Astronomy, Stockholm University \\
AlbaNova University Centre, SE-106 91 Stockholm, Sweden \\
\email{jorrit.leenaarts@astro.su.se}}

\date{Received: date / Accepted: date}

\maketitle


\begin{abstract}
Nearly all energy generated by fusion in the solar core is ultimately radiated away into space in the solar atmosphere, while the remaining energy is carried away in the form of neutrinos. The exchange of energy between the solar gas and the radiation field is thus an essential ingredient of atmospheric modeling. The equations describing these interactions are known, but their solution is so computationally expensive that they can only be solved in approximate form in multi-dimensional radiation-MHD modeling. In this review, I discuss the most commonly used approximations for energy exchange between gas and radiation in the photosphere, chromosphere, and corona.
\end{abstract}

\setcounter{tocdepth}{3}
\tableofcontents


\section{Introduction}
\label{sec:introduction}

The interaction of matter and radiation is an indispensable ingredient of solar and stellar atmospheric modeling. Around 98\% of the energy generated in the solar core is transported outwards first by radiative diffusion, then convection, and ultimately escapes into space in the photosphere of the Sun where the overlying solar material becomes transparent. The remainder of the energy escapes the Sun in the form of neutrinos
\citep{2011RPPh...74h6901T}.
The layers above the photosphere (chromosphere and corona) are hotter than radiative equilibrium models predict. Therefore deposition of non-thermal energy and conversion into heat must occur in these layers. The radiative energy losses \citep[ignoring the $\sim 10^{-6}$ fraction of energy carried away by the solar wind,][]{2012SoPh..279..197L} from the chromosphere and corona must balance this energy deposition in a time and space-averaged sense. Radiation is thus an essential ingredient in setting the structure of the outer solar atmosphere and the response of the atmosphere to non-thermal energy deposition. 

Modeling the interaction of radiation and matter in the solar atmosphere is in general a difficult problem. The specific intensity, the fundamental quantity used to characterise the radiation field, depends on seven parameters: three space dimensions, two angles describing direction, frequency, and time. In addition, the radiative transfer problem is non-linear and non-local: local changes to the intensity through absorption and emission depend on the intensity itself, and radiation emitted at one place in the atmosphere can influence other locations.

Writing general equations that describe the interaction of radiation and matter is not so difficult. Because of limitations in computing time these equations have so far been solved rather completely in one-dimensional geometry only. The Radyn code
\citep{1992ApJ...397L..59C,2002ApJ...572..626C},
is perhaps the most well-known example, but other codes are used too
\citep[e.g., Flarix, see][]{2009A&A...499..923K}.

However, solving these general equations in 2D and 3D is still beyond current computational capabilities. This is problematic because modeling convection requires at least 2D geometry, and fully modeling the rich physics caused by the interaction of the magnetic field with matter  requires 3D geometry. 

Since the pioneering simulations of 
\citet{1982A&A...107....1N},
a plethora of codes that aim to model solar and/or stellar atmospheres in 3D have been developed. The ones that I am aware of are: \textsc{Stagger} \citep[e.g.,][but many different versions of this code exist]{1998ApJ...499..914S},
\textsc{MURaM} \citep{2004PhDT-voegler,2017ApJ...834...10R},
\textsc{Bifrost} \citep{2011A&A...531A.154G},
COBOLD \citep{2012JCoPh.231..919F},
\textsc{Stellarbox} \citep{2015arXiv150707999W},
\textsc{RAMENS} \citep{2015ApJ...812L..30I,2016PhDT.........5I},
\textsc{MANCHA3D} \citep{2017A&A...604A..66K,2018A&A...618A..87K},
{\textsc{ANTARES} \citep{2017Ap&SS.362..181L},}
 and RADMHD \citep{2012SoPh..277....3A}. The latter is the only code that does not use the radiation treatment developed by \citet{1982A&A...107....1N}, and instead uses a simpler but much faster method.

Radiation-hydrodynamics in the solar atmosphere is a vast topic. In this review I limit myself to describing only the most commonly used approximations for radiative transfer in the photosphere, chromosphere, and corona in these multi-dimensional radiation-MHD codes. I do not discuss results obtained with any of these codes. An excellent review of solar magnetoconvection as studied using radiation-MHD simulations is given in \citet{2012LRSP....9....4S}. The field of radiation-MHD modeling of the combined photosphere, chromosphere, transition region, and/or corona has developed tremendously during the last 15 years. To my knowledge no recent review covering this development exists. Example starting points for studying applications are
\citet{2016A&A...585A...4C,2017Sci...356.1269M, 2018A&A...618A..87K}, and \citet{2019NatAs...3..160C}.

In the convection zone the radiation diffusion approximation holds to a high degree of precision and the energy exchange between radiation and the solar gas is close to zero. In the photosphere the gas loses large amounts of energy in the form of radiation which then largely escapes into space. The LTE assumption for the source function and opacity is still not too bad and even the grey approximation is still reasonably accurate. I devote a large fraction of this review discussing approximations for the photosphere and low chromosphere in Sect.~\ref{sec:multigroup}.

The situation in the chromosphere and transition region is more complex. The chromosphere is optically thin for optical continuum radiation and most radiative energy exchange takes place in a few strong spectral lines. Radiation scattering is important so that non-LTE effects must be taken into account, the ionisation balance of hydrogen and helium is out of equilibrium so that assuming LTE or statistical equilibrium to compute opacities or source functions is in general no longer accurate. Modeling energy exchange taking into account these complexities is discussed in Sections~\ref{sec:chromosphere} and~\ref{sec:noneq-ionisation}.

In the corona the physics of radiative energy losses and gains becomes again somewhat less complex. {Most coronal structures are} optically thin for all frequencies except in the radio regime. The radio regime lies in the far tail of the Planck function and does not contribute significantly to the radiative losses. For all other frequencies it  can be assumed that photon absorption does not take place and that all  photons emitted by the gas escape from the corona. It is discussed in Sect.~\ref{sec:corona}.


\section{Fundamentals}
\label{sec:fundamentals}

Excellent books that review the fundamentals of radiation hydrodynamics and modeling of stellar atmospheres are
\citet{1984oup..book.....M} and \citet{2014tsa..book.....H}. The first book focusses on fundamental theory, while the second one discusses modeling of 1D static and moving atmospheres. Below I briefly touch upon some general aspects that are relevant for the methods discussed in Sect.~\ref{sec:multigroup}\,--\,\ref{sec:new-developments}.

\subsection{The MHD equations including radiation}
The dominant paradigm for multidimensional modeling of the solar atmosphere has been the magnetohydrodynamics (MHD) approximation. The MHD equations for the density, momentum and internal energy including radiation terms can be written as
\bea
 \frac{\partial \rho }{\partial t} & = & - \grad \cdot \left( \rho \vv \right), \\
  \frac{\partial \vp }{\partial t} & = & - \grad \cdot \left( \vv \otimes \vp - \vec{\tau} \right)  - \grad P + \vJ \times \vB + \rho \vg - \grad \vP_\mathrm{rad},  \\
   \frac{\partial e}{\partial t}  & = & - \grad \cdot \left( e \vv \right)  - P \grad \cdot \vv + Q + \Qrad,
\eea
with $\rho$ the mass density, $\vv$ the velocity vector, $\vp$ the momentum density, $\vec{\tau}$ the stress tensor, $P$ the gas pressure, $\vJ$ the current density, $\vB$ the magnetic field vector, $\vg$ the acceleration due to gravity, $\vP_\mathrm{rad}$ the radiation pressure tensor, $e$ the internal energy, $\Qrad$ the heating or cooling owing to radiation, while $Q$ expresses energy exchange by any other processes such as dissipation of currents, heat conduction, and viscosity. 

The assumptions under which MHD is valid tend to be fulfilled in the photosphere and convection zone as well as the corona, but break down in the chromosphere and transition region where the frequencies of collisions between particles become smaller than the cyclotron frequencies of ions and electrons.
\citep[e.g.,][]{2014PhPl...21i2901K}.

Several radiation-MHD codes have been extended to include some effects beyond MHD, through including the ambipolar diffusion term, the Hall term, and/or Biermann's battery term, to the induction equation
\citep{2012ApJ...753..161M,2012ApJ...750....6C,2017A&A...604A..66K,2018A&A...618A..87K}
These inclusions retain the single fluid description. This greatly simplifies the treatment of the radiative term $\Qrad$, because energy lost or gained by the gas modifies the internal energy of the gas as a single entity, instead of modifying the energy of the electrons and different species of atoms, ions, and molecules separately.

Efforts are underway to move beyond single-fluid {radiation-}MHD to a multi-fluid description, treating neutrals, ions, and/or electrons as separate fluids each with their own temperatures and velocities. Radiative transitions can modify the internal energy and change the ionisation state of atoms, and in a multi-fluid description these must be computed in detail
\citep[e.g.,][]{2014PhPl...21i2901K}.
This review {does not} discuss radiation-hydrodynamics in multi-fluid models.


\subsection{Energy density of radiation and matter}

\begin{table}
\caption{Comparison of energy densities and radiative flux}
\label{table:edens-comparison}       
\begin{tabular}{llll}
\hline\noalign{\smallskip}
 &$E_\rad$ (J m$^{-3})$ & $e$ (J m$^{-3}$)  & $E_\rad/e$ \\
photosphere & $0.84$ & $1.2\times10^4$ &  $7.0\times10^{-5}$\\
chromosphere & 0.37 &  $17$& $2.2\times10^{-2}$ \\
corona & 0.22 & $0.21$& $1.1$ \\
\noalign{\smallskip}\hline
\end{tabular}
\end{table}

It is useful to compare the energy density and flux of the radiation field to the energy density of the gas in the solar {atmosphere}. To get an estimate we assume that the radiation field is isotropic and given by the Planck function at $T_\rad=5777$~K. Then the energy density is 
\be
E_\rad = \frac{4\sigma}{c} T_\rad^4 =  0.84\ \mathrm{J\, m}^{-3}, \label{eq:fluxinatmos}
\ee
with $\sigma$ the Stefan--Boltzmann constant. Assuming that the solar surface is a radiating blackbody at the same temperature gives the radiative flux at the surface as
\be
F_\rad = \sigma T_\rad^4 = 6.3\times10^7 \ \mathrm{W\, m}^{-2} ,
\ee
Table~\ref{table:edens-comparison} compares the radiative energy density to the internal energy density $e$ of the solar gas for typical photospheric, chromospheric and coronal values assuming the solar gas has an ideal gas equation of state ($e=n k_\mathrm{B} T$). The radiative energy density in the chromosphere and corona is corrected for the non-isotropy of the radiation above the photosphere. The energy density of the radiation is much lower than the energy density of the gas in the photosphere and chromospere, but in the corona they are about equal. The corona is however optically thin for most photospheric radiation and only little absorption or scattering of radiative energy by the gas occurs.


\subsection{Radiation pressure and force} \label{subsec:radforce}

The force exerted by the radiation pressure tensor is typically ignored under normal solar conditions because it is small compared to other forces. To illustrate this one can compare the radiation pressure of isotropic black body radiation to the gas pressure in the photosphere. This pressure is
\be
\frac{4 \sigma}{3c} T^4 = 0.27~\mathrm{Pa}
\ee
at a photospheric temperature of $T=5700$ K, while the gas pressure in the photosphere is roughly $10^4$~Pa. Similarly, one can compute a rough estimate of the upward acceleration of photospheric material by radiation:
\be
a_\mathrm{rad} \approx \frac{\kappa F}{c} = 3 \times 10^{-3}~\mathrm{m\,s}^{-2},
\ee
with $\kappa$ the Rosseland opacity per mass unit and $F$ the frequency-integrated radiative flux, both taken in the photosphere.  The radiative acceleration is much smaller than the downward-directed solar surface gravity acceleration $g=274$~m\,s$^{-2}$. 


\subsection{Energy exchange between radiation and matter}

In absence of an absorbing medium, the monochromatic radiative flux divergence $\divF_\nu$\ is zero. If a medium (in the solar case a gas or plasma) is present, then the rate per volume with which the material gains energy from the radiation field is given by
\be
Q_\rad =  - \int_0^\infty \divF_\nu \, \dnu =  -\divF,
\ee
with \vF\ the total radiative flux. The total radiative flux is an integral over frequency, and as far as the internal energy of the gas is  concerned, the exact distribution of the flux over frequency is not important. Typically, radiation-MHD simulations of the solar atmosphere aim to reproduce the correct detailed behaviour of the gas only. The computation of the radiation field only has to reproduce the correct heating and cooling, but does not need to reproduce the correct spectral energy distribution. This allows for {large} simplifications in treating the radiation without sacrificing too much accuracy in the value of the total flux divergence.


\subsection{Explicit expression of the radiative flux divergence}

The monochromatic flux is defined in terms of the intensity \Inu\ as
\be
 \vF_\nu = \oint  \vn \Inu \,\dd\Omega, \label{eq:fluxdef}
\ee
with $\vn$ the unit vector pointing in the direction of $\Omega$. Substitution of this equation into the integral over all directions of Eq.~\eqref{eq:transfer_form1}  yields an expression of the total radiative flux divergence in terms of emissivity or source function $\Snu=\jnu/(\kappa_\nu \rho)$, intensity and opacity:
\bea
\divF &=&  \int_0^\infty  \oint \left(\jnu - \knu \rho\Inu \right) \, \dd \Omega \,\dnu \label{eq:generalfluxdiv} \\ 
&=&   \int_0^\infty  \oint \knu \rho \left(\Snu - \Inu \right) \, \dd \Omega \,\dnu, \nonumber
\eea
with $\knu$ the extinction coefficient per unit mass. If one assumes that both the emissivity and extinction coefficient do not depend on direction then Eqs.~\eqref{eq:generalfluxdiv} reduce to
\be
\divF =  \int_0^\infty 4 \pi \knu \rho \left(\Snu - \Jnu \right) \, \dnu, \label{eq:novelfluxdiv}
\ee
where 
\be
\Jnu= \frac{1}{4 \pi} \oint  \Inu \,\dd\Omega \label{eq:JfromI}
\ee
is the angle-averaged {intensity}.

This assumption is reasonable for bound-free transitions and other continuous processes. For bound-bound transitions it is not valid when flow velocities are of the same order or larger than the {thermal speed} $\sqrt{2kT/m+v_\mathrm{turb}^2}$, with $m$ the mass of the atom or ion that is involved, and $v_\mathrm{turb}$ the microturbulent velocity. Nevertheless, the assumption is almost always taken to be valid because it allows {large} simplifications.


\subsection{Light travel time and hydrodynamical timescales}

Typical bulk flow velocities in the solar atmosphere range from 1~\kms\ in convective upflows to $>300$\,\kms\ in chromospheric evaporation following solar flares
\citep{2015ApJ...807L..22G}.
Alfv\'en velocities up to $2.2\times10^3$~\kms\ have been measured in the solar corona by studying properties of coronal loop oscillations 
\citep[e.g.,][]{1999ApJ...520..880A,2018ApJ...860...31P}.
These values are much lower than the speed of light.

The typical hydrodynamic timescale in the photosphere is $\sim5$~minutes, it is $\sim1$~minute in the chromosphere and a few minutes in the corona. The light crossing time is of the order of a second even for the largest coronal structures.

Because of the low flow speeds and long hydrodynamic timescales compared to light crossing times, it is typically assumed that light travel times can be ignored in solar radiation hydrodynamics. That means that the computation of the radiative flux divergence at a given time $t$ only depends on the state of the atmosphere at time $t$. The time-dependent transfer equation for the intensity in direction \vn\ then simplifies from
\be
\frac{1}{c} \frac{\partial I_\nu}{\partial t} + \vn \cdot \nabla  \Inu = \jnu - \anu \Inu
\ee
to
\be
 \vn \cdot \nabla  \Inu  = \jnu - \anu \Inu, \label{eq:transfer_form1}
\ee
with $\jnu$ the emissivity and $\anu = \knu \rho$ the extinction coefficient per volume. In other words, it is assumed that the radiation field adapts instantaneously to changes in the state of the solar gas.

\subsection{Diffusion approximation}
At very large optical depth the following approximation holds for  $\Jnu$:
\be
\Jnu \approx \Snu +\frac{1}{3} \frac{\dd^2 \Snu}{\dd \tnu^2}, \label{eq:Japprox}
\ee
where the second derivative should be taken along the direction of the gradient of $\Snu$\footnote{formally, $\dd^2 \Snu / \dd \tnu^2 = 1/(\knu \rho)^2 \vec{n}^\mathrm{T} H \vec{n}$, with $\vec{n} = \nabla \Snu / \left|\nabla \Snu \right|$ and $H$ the Hessian matrix of $\Snu$.}.
For the flux the following approximation holds:
\be
\Fnu \approx - \frac{4 \pi}{3 \knu \rho} \nabla \Snu. \label{eq:Fapprox}
\ee
At depths where the optical depth at all frequencies is much larger than unity, one can derive the diffusion approximation for the total radiative flux  from from Eq.~\eqref{eq:Fapprox}  if one also assumes that the source function is equal to the Planck function \Bnu:
\be
\vF  \approx - \frac{16}{3} \frac{\sigma T^3}{\kR \rho} \nabla T,
\ee
with the Rosseland opacity defined as
\be
\frac{1}{\kR} =   \left( \int_0^\infty \frac{1}{\knu} \frac{\dd \Bnu}{\dd T} \, \dd \nu \right) \left/  \left( \int_0^\infty \frac{\dd \Bnu}{\dd T} \, \dnu \right) \right..
\ee

Methods to compute the total flux divergence that rely on the solution of the radiative transfer equation must converge to the diffusion approximation at large depth.  Methods that require a numerical approximation of the source function must use one of at least second order in order to recover Eq.~\eqref{eq:Japprox}.


\section{Radiative transfer in the photosphere: multi-group radiative transfer}
\label{sec:multigroup}

Multi-group radiative transfer was first introduced in 
\citet{1982A&A...107....1N}.
It is a method that approximates the frequency integral in Eq.~\eqref{eq:novelfluxdiv} by a sum over a limited number $N$ so called radiation groups (or radiation bins):
\be
\int_0^\infty 4 \pi \kappa \rho \left(\Snu - \Jnu \right) \, \dnu \approx \sum_{i=1}^N 4 \pi \kappa_i \rho \left( S_i - J_i \right), \label{eq:approxfluxdiv}
\ee
with $\kappa_i$, $S_i$, and $J_i$ the opacity, source function and radiation field in each group, as defined below.

The rationale of the method is the realisation that the $\Lambda$-operator that produces the angle-averaged radiation field from the source function depends on the opacity only, and is a linear operator. While it is customary to write $\Lambda_\nu[\ldots]$ because the opacity changes with frequency, a better notation would be $\Lambda_{\kappa}[\ldots]$.
For two frequencies $\nu_1$ and $\nu_2$ with identical opacities everywhere in the atmosphere $\kappa_1 = \kappa_2 = \kappa$ one can thus write
\be
 J_{\nu_1} + J_{\nu_2} =  \Lambda_{\nu_1} [S_{\nu_1}] +  \Lambda_{\nu_2} [S_{\nu_2}] = \Lambda_\kappa [S_{\nu_1}+S_{\nu_2}] \label{eq:linearlambda}
\ee
In order to derive the approximation in Eq.~\eqref{eq:approxfluxdiv} one first approximates the frequency integral with a sum over discrete frequencies $\nu_j$ with summation weights $w_j$. These frequencies are then grouped
into $N$ bins, each labeled $i$, that have similar opacities. Finally, Eq.~\eqref{eq:linearlambda} is used to define a bin-integrated source function:
\bea
\int_0^\infty  \kappa \left(\Snu - \Jnu \right) \, \dnu &  \simeq & \sum_j w_j \kappa_j \left( S_j - J_j \right) \\ \label{nordlund1}
= \sum_{i=1}^N\sum_{j(i)} w_j \kappa_j \left( S_j -J_j \right) &= &  \sum_{i=1}^N \sum_{j(i)} w_j \kappa_j  \left(S_j - \Lambda_j[S_j] \right) \\
\approx  \sum_{i=1}^N \kappa_i \left( \sum_{j(i)} w_j S_j  - \Lambda_i[\sum_{j(i)} w_j S_j ] \right) &\equiv& \sum_{i=1}^N \kappa_i \left( S_i - \Lambda_i[S_i] \right) \\
&\equiv& \sum_{i=1}^N \kappa_i \left(S_i -J_i \right). \label{nordlund4}
\eea
Here $j(i)$ is the set of all frequencies with similar opacities that are placed in bin $i$. The bin-integrated source function is constructed from the frequency-dependent source function using
\be
S_i = \sum_{j(i)} w_j S_j
\ee

What is left is now is to define how to group frequency points in the various bins, how to define the representative opacity for each bin $\kappa_i$, and how to define the frequency-dependent source function $S_j$. The opacity and source function are in general functions of at least frequency, temperature and density. Additional dependencies are on the velocity field in the atmosphere, the ray direction, possible location-dependent abundances, and in the case of non-LTE radiative transfer, on the radiation field. 

Section~\ref{subsec:group-opacities} discusses how to sort frequencies into opacity groups. In Sect.~\ref{subsec:MGRTL} it is assumed that the source function is given by the Planck function. This assumption is relaxed to allow for coherent scattering in Sect.~\ref{subsec:MGRTS}.

\subsection{Sorting frequencies into groups} \label{subsec:group-opacities}

\begin{figure*}
\centering
 \includegraphics[width=0.75\textwidth]{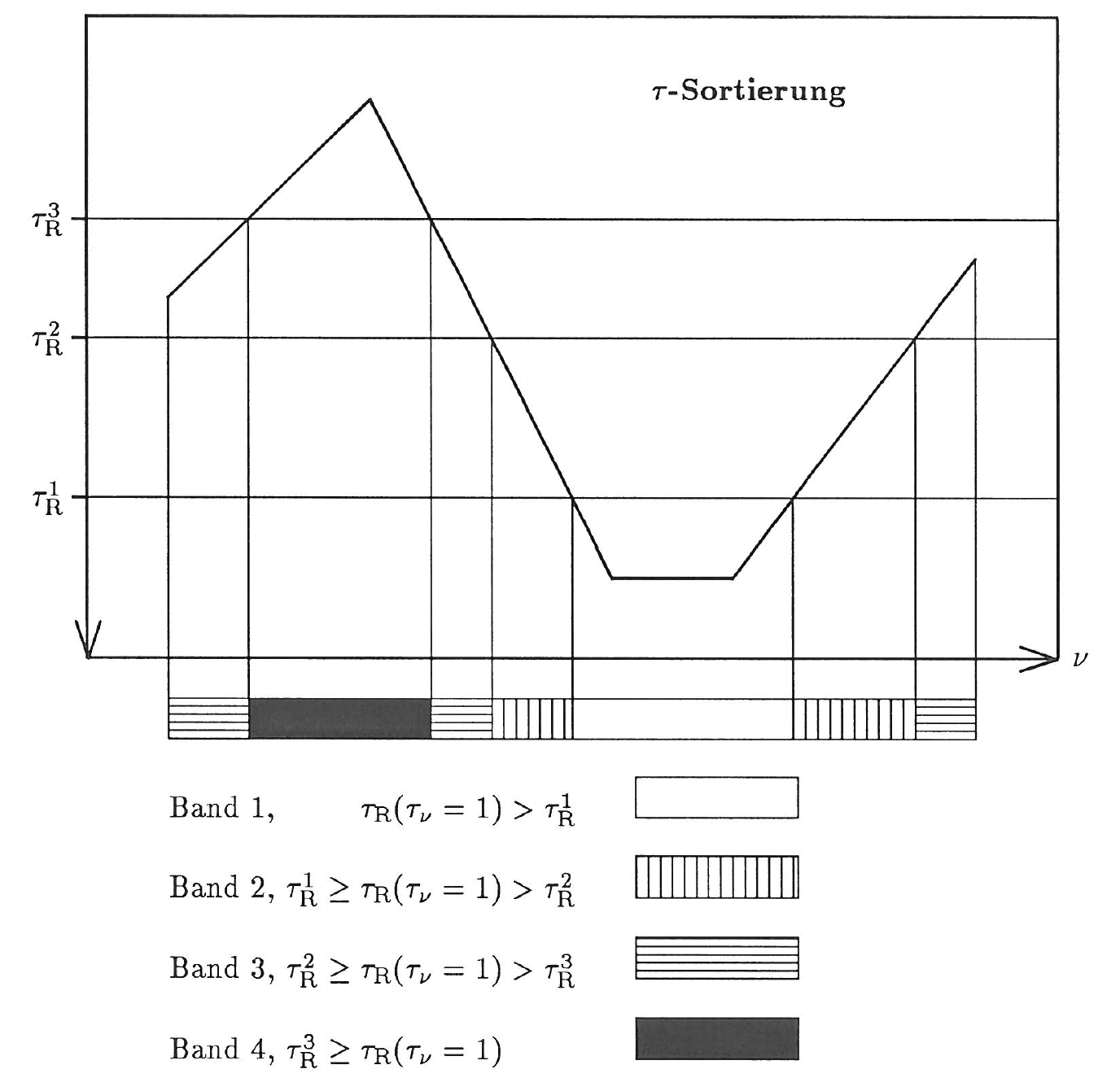}
\caption{Illustration of the principle of $\tau$-sorting. The horizontal axis is frequency, while the vertical axis shows the Rosseland optical depth in the 1D reference atmosphere. The solid line is the Rosseland optical depth where the monochromatic optical depth unity is reached. The frequencies are divided into four bins using the three border values $\tR^1$\,--\,$\tR^3$. Adapted with permission from \citet{1992PhDT-ludwig}.}
\label{fig:tau-sorting-ludwig}       
\end{figure*}

Grouping frequencies and defining a bin opacity are necessarily somewhat crude. The density and thus opacities drop roughly exponentially with height, and opacities at a given point in the atmosphere vary over many orders of magnitude with frequency. The aim of the multi-group opacities is to approximate the full radiative transfer with sensitivity spanning from the low-opacity continua that form in the photosphere to strong lines that form in the mid and ideally even the upper chromosphere. 

The de-facto standard method for grouping opacities is called $\tau$-sorting: First a 1D reference atmosphere is chosen, for which the opacities and vertical optical depth scales at all frequency points $j$ are computed. The atmosphere is then divided into various height bins.  Frequencies at which optical depth unity is reached in height bin $i$ are assigned to radiation bin $i$. This height sorting can in principle be done directly on a geometrical height scale, but most radiation-MHD codes follow
\citet{1982A&A...107....1N}.
They set the borders between the height bins in terms of the Rosseland optical depth scale $\tR$: Let the borders be defined at a number of Rosseland optical depths $\tR^k$. A frequency $\nu_j$ then belongs to bin $i$ if
\be
 \tR^{k-1}<  \tR(\tau_\nu =1) \leq  \tR^k.
\ee
A common choice is to put the border values $\tR^k$ equidistantly spaced in $\log{\tau_{\mathrm{R}}}$. The method is illustrated in Fig.~\ref{fig:tau-sorting-ludwig}.

Opacities are {generally} well-approximated by their LTE values in the photosphere.
LTE opacities in a static atmosphere are functions of temperature, density and elemental composition only, but even then it requires a large effort to accurately compute them. For historical reasons this is commonly done using an intermediate step called Opacity Distribution Functions (ODFs). ODFs were developed originally to speed up computations used in modeling of 1D LTE radiative equilibrium stellar atmospheres
\citep[e.g.,][]{1975A&A....42..407G,1979ApJS...40....1K}.

\begin{figure*}
\centering
 \includegraphics[width=0.49\textwidth]{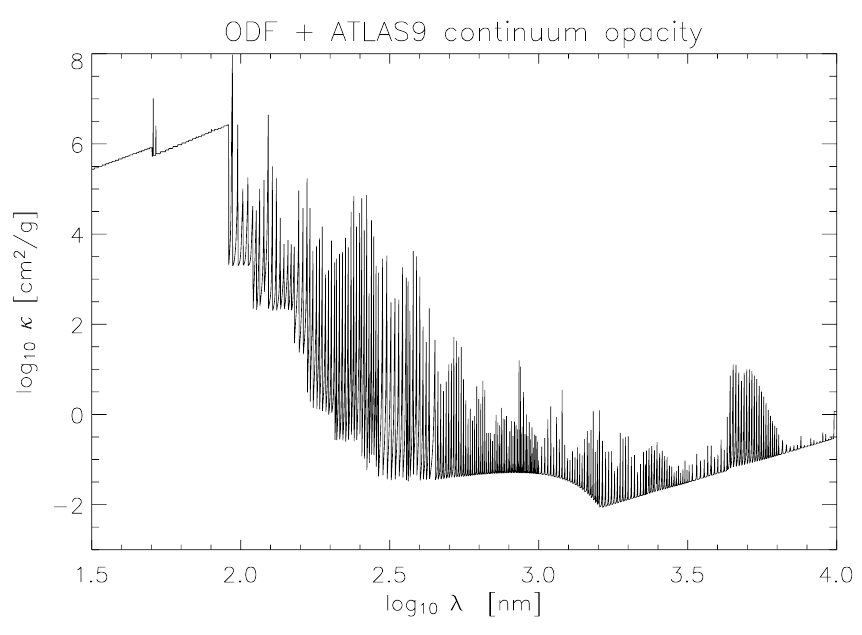}
 \includegraphics[width=0.49\textwidth]{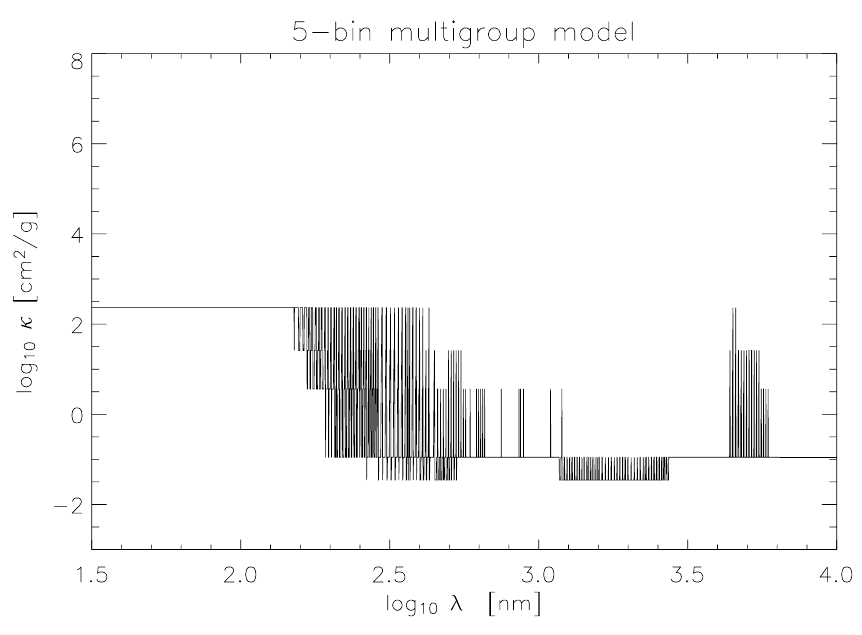}
\caption{Illustration of the concept of group mean opacity. Left-hand panel: Monochromatic extinction coefficients based on Opacity Distribution Functions  for $T=4470~K$ and $ P = 1.3\times10^4$~Pa, corresponding roughly to the upper photosphere. 
Right-hand panel: The opacity that is ``assigned'' to each wavelength using a 5-group scheme. The large continous variation in the left-hand panel is replaced by only five discrete opacities. Note that the high monochromatic opacities in the UV are replaced by a much lower group opacity.
Adapted with permission from \citet{2004A&A...421..741V}, copyright by ESO.}
\label{fig:voegler2004-figs2-3}       
\end{figure*}

The method used to compute an appropriate mean opacity in a bin $i$ from the opacities $\kappa_j$ depend on whether one assumes an LTE or non-LTE source function and are described in Sects.~\ref{subsec:MGRTL} and~\ref{subsec:MGRTS}. An illustrative solution assuming an LTE source function is given in Fig~\ref{fig:voegler2004-figs2-3}.

\subsection{Multi group radiative transfer with LTE source function} \label{subsec:MGRTL}

The source function in the solar atmosphere is in general not equal to the Planck function because at some height in the atmosphere densities become too low to set up Saha--Boltzmann populations through collisions. However, detailed non-LTE computations in 1D models show that in the {deep} photosphere (roughly defined here as $-100~\mathrm{km} <z< 200~\mathrm{km}$, with the $z=0$ defined as the location where $\tau_{500~\mathrm{nm}}=1$), the source function is almost exactly equal to the Planck function
\citep[see Fig.~36 of][]{1981ApJS...45..635V},
and the opacities are close to their LTE values. \edt{}At larger heights this is no longer the case for lines and continua of neutral atoms with a low ionisation potential because they tend to be over-ionized by the radiation from below.
Nevertheless, one can expect that assuming LTE for both the source function and the opacity is a good approximation for computing the flux divergence in the photosphere. The source function in group $i$ is then given by
\be
S_i =  \sum_{j(i)} w_j B_j.
\ee
This expression only depends on temperature and can thus easily be precomputed and stored in a 1D lookup table.

Once the opacities $\kappa_j$ are grouped into the $N$ bins, one still needs to compute an appropriate bin opacity $\kappa_i$. Choosing the numerical equivalent of the Rosseland opacity in each bin ensures that the diffusion approximation is recovered deep in the atmosphere:
\be
\frac{1}{\kappa_i^\mathrm{R}} =   \left( \sum_{j(i)} w_j  \frac{1}{\kappa_j} \frac{\dd B_j}{\dd T}  \right) \left/ \left(  \sum_{j(i)} w_j  \frac{\dd B_j}{\dd T}  \right) \right. .
\ee

This choice of bin opacity is however not appropriate at low optical depths, where photons are mainly in the free streaming regime. Following approximations valid at small optical depth put forward in 
\citet{1970stat.book.....M},
and further developed in 
\citet{1992PhDT-ludwig},
it turns out that the Planck-mean opacity \kB\ is a good choice for the outermost layers of the atmosphere:
\be
\kappa_{i}^B =  \frac{\sum_{j(i)} w_j \kappa_j B_j}{ \sum_{j(i)}  w_j B_j }.
\ee
Somewhere in the atmosphere one should make a transition from one opacity definition to the other. This can be achieved through defining the group opacity as
\bea
\kappa_i & = &W \kappa_i^B + (1-W) \kappa_i^\mathrm{R}, \\
W & = & \mathrm{e}^{-\tau_i/\tau_0} .
\eea
where $\tau_0$ is an adjustable parameter of order unity and $\tau_i$ the vertical optical depth. Computing $\tau_i$ in the MHD simulation is somewhat computationally expensive and instead it is estimated, for example through using the relation between mass density and optical depth in the 1D reference atmosphere used for the sorting of frequencies into bins
\citep{1992PhDT-ludwig,2004A&A...421..741V}.

Another option is to base the transition on the local mean free path:
\be
W = \mathrm{e}^{-l_i/(\kappa_i \rho)},
\ee
where $l_i$ is a typical length scale over which the bin-integrated source function changes
\citep{2000ApJ...536..465S}.
The advantage of this method is that it does not depend on the properties of a 1D reference model.

For a fixed elemental composition, $\kappa_i$ depends only on a combination of any two thermodynamic parameters (for example $e$ and $\rho$). Like the group-integrated source function, they are commonly precomputed and put in a 2D lookup table.

Extensive discussions of multi-group radiative transfer assuming an LTE source function can be found in the PhD-thesis of
\citet[][in German]{1992PhDT-ludwig}
and in 
\citet{2004A&A...421..741V}.

\subsection{Multi group radiative transfer with non-LTE source function} \label{subsec:MGRTS}

The assumption of LTE for the source function is accurate in the photosphere, but breaks down in the chromosphere and transition region. Here the energy exchange is dominated by the resonance lines of \HI, \CaII, \MgII, and \HeII\
\citep[see for example Fig. 49 of][]{1981ApJS...45..635V}.
These lines can have photon destruction probabilities
\be
\epsilon = \frac{C_{ul}}{A_{ul} + C_{ul} + \Bnu B_{ul}} < 10^{-4},
\ee
with $C_{ul}$ the downward collisional rate coefficient and $A_{ul}$ and $B_{ul}$ Einstein coefficients for spontaneous and induced deexcitation. Scattering should therefore be taken into account.

Scattering has a strong damping effect on the amplitude of $\divF_\nu/\rho$. In LTE it is given by
\be
\frac{\divF_\nu^\mathrm{LTE}}{\rho} = 4 \pi \knu  \left( \Bnu - \Jnu \right).
\ee
If one assumes the presence of a coherently scattering line {in} a two-level atom at frequency $\nu$, {then the source function becomes 
\be
\Snu = (1-\epsilon) \jnu + \epsilon \Bnu.
\ee
In that case} the flux divergence per mass unit is 
\be
\frac{\divF_\nu^\mathrm{NLTE}}{\rho}  = 4 \pi \epsilon \knu  \left( \Bnu - \Jnu \right),
\ee
so that for a given difference between \Bnu\ and \Jnu, the amplitude of the radiative energy exchange in non-LTE can be orders of magnitude smaller than in LTE. 

\citet{2000ApJ...536..465S}
extended the method presented in Sect.~\ref{subsec:MGRTL} to include coherent two-level scattering, as an approximation to the much more complicated full non-LTE radiative transfer problem. He assumed that the monochromatic extinction coefficient can still be computed in LTE. This assumption is rather accurate for the resonance lines of \HI, \CaII, \MgII, whose lower levels are the ground state of a dominant ionisation state. In addition he assumed that the scattering coefficient in a spectral line is given using the approximation by 
\citet{1962ApJ...136..906V}, 
so that it is independent of the actual line, and only depends on frequency, temperature and electron density. 
Starting from the monochromatic two level source function  $\Snu = (1-\epsilon_\nu) \Jnu + \epsilon_\nu \Bnu$ and following a reasoning similar to the one given in Sect.~\ref{subsec:MGRTL}, he arrives at an expression for the group source function:
\be
S_i = \epsilon_i J_i + t_i. \label{eq:skartlien_source}
\ee
Here $\epsilon_i$ represents a group-averaged scattering probability, and $t_i$ the group-integrated thermal production of photons. He also derives an expression for the group extinction coefficient $\kappa_i$.  Similar to the LTE case, $\epsilon_i$, $t_i$, and $\kappa_i$ have different expressions for the diffusion regime and in the free streaming regime in the outer atmosphere. An important difference is that the streaming quantities now depend on the monochromatic mean intensity \Jnu\ in a 1D reference atmosphere. This means that the quantities must be recalculated for each different target of simulations (e.g., sunspots or quiet Sun).  Simulations containing a variety of structures might suffer from inaccuracies because a 1D atmosphere cannot be representative of all atmospheric structures.

Another difference compared to the method employing an LTE source function is that the computation of \Qrad\ now requires solution of Eq.~\eqref{eq:skartlien_source} together with the transfer equation because $J_i= \Lambda[S_i]$. This is typically done through accelerated $\Lambda$-iteration. Because of the multidimensional geometry, a local approximate $\Lambda$-operator 
\citep[which is equivalent to Jacobi-iteration, see][]{1986JQSRT..35..431O}
is efficient and easily coded, such as in the Oslo \textsc{Stagger} Code
\citep{2000ApJ...536..465S}. 
Gauss--Seidel iteration
\citep{1995ApJ...455..646T}
 offers superior convergence speed and has been implemented in the \textsc{Bifrost} code
 \citep{2010A&A...517A..49H,2011A&A...531A.154G}.

 \subsection{Solving the transfer equation}

\begin{figure}
\centering
 \includegraphics[width=8.8cm]{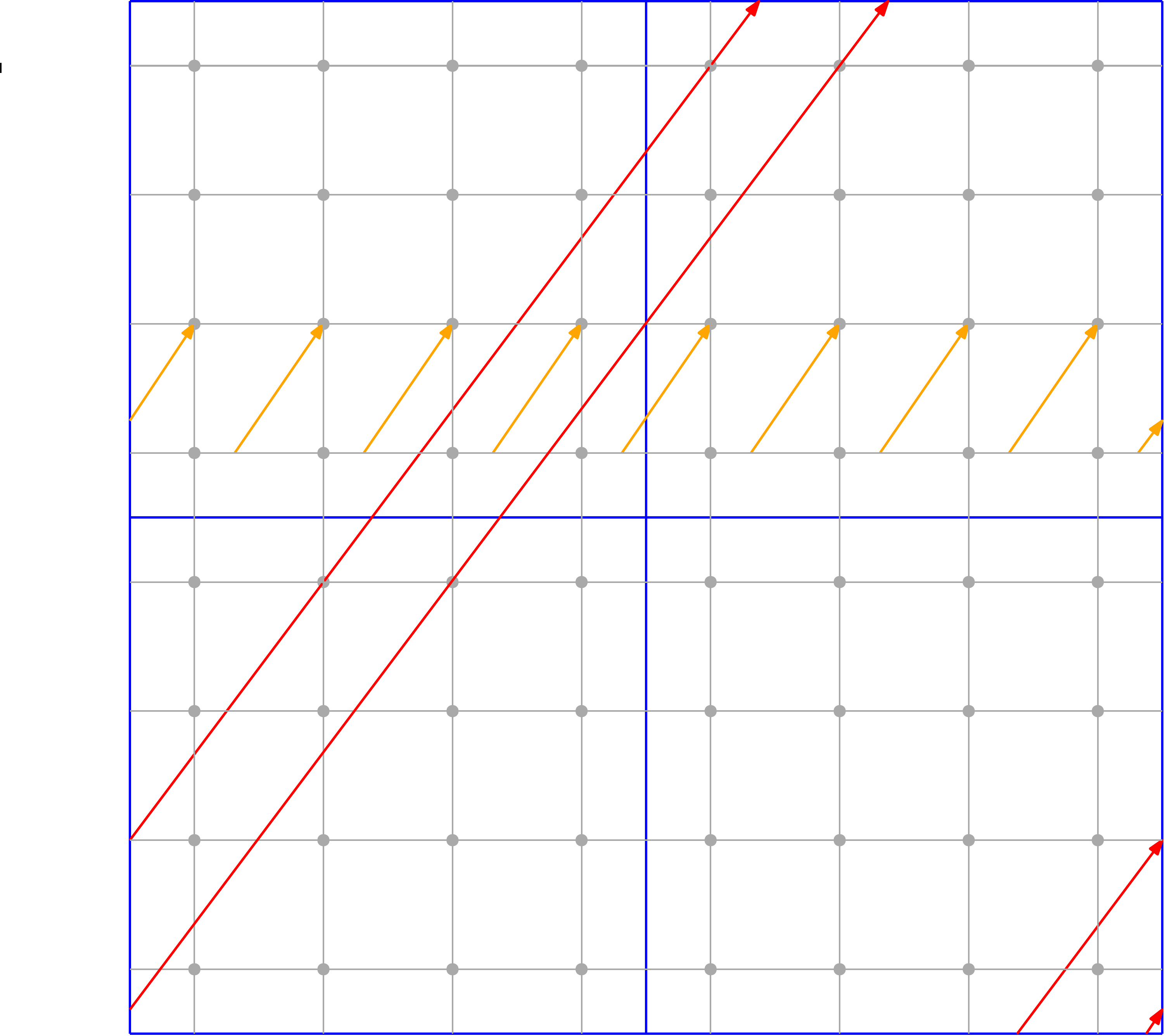}
\caption{Illustration of short and long characteristics used for solving the transfer equation in 2D decomposed domains. Grey dots indicate grid points where the intensity should be computed, with grey lines connecting those grid points. Blue lines indicate subdomain boundaries, where it is assumed that the horizontal boundaries are periodic. Red arrows indicate two examples of long characteristics, while the orange arrows indicate short characteristics. Note that the long characteristics cross multiple subdomain boundaries and wrap around the periodic horizontal boundary Each subdomain contains a piece of the long characteristic of variable length.}
\label{fig:longchar-shortchar}       
\end{figure}

Most modern radiation-MHD codes are parallelized to make use of large supercomputers with distributed memory. Typically, the simulated domain is represented on a 3D Cartesian grid. The domain is split into smaller subdomains and each CPU handles the required computations on its own subdomain, communicating information to other subdomains as needed. This architecture scales well for the MHD equations: they are local and require communication with neighbouring subdomains only. 

Radiation is however intrinsically non-local. An emitted photon can traverse many subdomains before undergoing another interaction with the solar gas, so communication between subdomains is not necessarily local. This problem is shared between radiation-MHD codes that aim to compute a reasonable approximation of the flux divergence, and non-LTE radiative transfer codes
such as Multi3d \citep{2009ASPC..415...87L} and PORTA \citep{2013A&A...557A.143S},
wich aim to accurately compute the emergent spectrum from a given model atmosphere. Consequently, there is a large amount of  literature addressing efficient solutions of the transfer equation in multidimensional geometries and/or decomposed domains
\citep[e.g.,][]{1988JQSRT..39...67K,2003ASPC..288..405A,2005A&A...429..335V,2006A&A...448..731H,2010A&A...517A..49H,2013A&A...549A.126I,2013A&A...557A.143S}. 
I will only briefly touch upon some aspects.

Solving for the flux $\vF$ or for the angle-averaged radiation field $J$ requires computation of the intensity for a number of different directions at all points on the numerical grid. In practice there are two methods that are used in radiation-MHD codes: {\it short characteristics} (SCs) and {\it  long characteristics} (LCs). Both are illustrated in Fig.~\ref{fig:longchar-shortchar}. 

Short characteristics solve the transfer equation along short line segments (orange in Fig.~\ref{fig:longchar-shortchar}), starting at a grid cell boundary and ending at a grid point where the intensity is desired. Along the line one typically computes a numerical approximation of the formal solution:
\be
I(\tau) = I(\tau=0) \, \rme^{-\tau} + \int_0^\tau S(t) \, \rme^{t-\tau} \dd t, \label{eq:formal_solution}
\ee
where the optical thickness scale has its zero point at the start of the SC and $\tau$ is the optical thickness along the entire SC. Intensities $I(\tau)$ are computed at the grid points (grey circles in Fig.~\ref{fig:longchar-shortchar}). Computation of $I(\tau=0)$, which is needed at the start of the orange arrow, thus requires interpolation from the grid points. SC methods are therefore somewhat diffusive and coherent beams of photons disperse. High-order interpolation schemes can alleviate, but not eliminate, this diffusion. In practice this diffusion is typically not a problem, given that the photosphere is emitting photons everywhere and both the source function and the resulting radiation field are rather smooth. 

The transfer equation is solved along all SCs in a sequential order, starting from a known boundary condition (the diffusion approximation at the bottom of the atmosphere for rays going up, and typically zero for SCs going down from the top of the atmosphere). The method was introduced for 2D cartesian geometry by 
\citet{1988JQSRT..39...67K}.
An in-depth description of the method in 3D Cartesian geometry is given in 
\citet{2013A&A...549A.126I}.
The ordered fashion in which SCs must be computed leads to complications with spatial domain decomposition. An example method of how to achieve reasonable parallelism despite this ordering can be found in
\citet{2013A&A...557A.143S}. 
Short characteristics can be easily computed along any angle. Typical ray quadratures (i.e., the set of angles chosen to numerically compute Eq.~\eqref{eq:fluxdef} or~\eqref{eq:JfromI}) that are in use are the angle sets from
\citet{carlson1963},
or equidistant in azimuth and using a Gaussian quadrature in inclination.

\citet{1999A&A...348..233B}
present a method to compute SCs on unstructured grids. It is not currently in use in the common radiation-MHD codes that use fixed Cartesian grids, but might be of great use for codes that use unstructured or adaptive grids.

The long characteristic method traces rays from the lower boundary of the domain to the upper boundary (red line segments in Fig.~\ref{fig:longchar-shortchar}). An LC does generally intersect only a few grid points. Interpolation of the source function and the opacity from the grid points to the LC and interpolation of the intensity along the LC back to the grid points is thus necessary. The transfer equation along an LC can be solved using the formal solution (Eq.~\eqref{eq:formal_solution}), or by solving the second-order form of the transfer equation
\citep{1964CR....258.3189F}.

Long characteristics allow photons to travel in a straight line and are thus not diffusive. Efficient parallel algorithms exist for solving along LCs in decomposed domains
\citep{2006A&A...448..731H}.
However, this parallel algorithm is only easily implemented when LCs cross cell boundaries exactly through grid points, which limits the application to  grids with fixed spacing and a maximum of 26 directions: both directions along three axes, six face diagonals, and 4 space diagonals in  a cubic grid
\citep[e.g.,][]{2019MNRAS.482L.107P}.
Arbitrary ray quadratures can be implemented at the expense of code simplicity.

Both the SC and LC method can handle exactly horizontal rays, but such rays require implicit solution methods in case of periodic boundary conditions in the horizontal direction, which is the default for solar atmosphere radiation-MHD simulations.  This adds additional coding complications in the usual case that parallelism is implemented through spatial domain composition. Horizonal ray directions are therefore usually avoided.

\subsection{Computation of the heating rate from the intensity and source function}

Analytically, the two expressions for the heating rate in a bin $Q_i=-\divF_i = 4\pi \kappa_i \rho_i (J_i-S_i)$ are equivalent. In case of actual numerical computation this is no longer the case. 

In the deep atmosphere the diffusion approximation holds. Eq.~\eqref{eq:Japprox} shows that while 
$S_i$ and $J_i$ increase with depth because of the increase in temperature, their difference becomes smaller, and at some point roundoff errors  become noticeable. These errors are then amplified by the exponential increase of the density with depth. Using the source function and radiation field to compute $Q_i$ is thus unstable in the deep layers. Instead, computing the flux from the intensity using the numerical equivalent of Eq.~\eqref{eq:fluxdef} and then taking the divergence is stable in the deep layers. Because the radiative flux is small compared to the energy density of the gas, an error in computation of its divergence will not lead to large errors in the internal energy.

In the upper layers the situation is reversed: the radiative flux is large compared to the energy density of the gas (see Table~\ref{table:edens-comparison} and Eq.~\eqref{eq:fluxinatmos}), and a small error in computation of the flux divergence from the intensity will lead to a large error in $Q_i$ and $e$. Using $4\pi \kappa_i \rho_i  (J_i-S_i)$ is stable however, because of the generally large split between $S_i$ and $J_i$.  

Following the suggestion by 
\citet{1999A&A...348..233B},
{most 3D codes that use short characteristics compute $Q_i$ using the flux divergence at large optical depth, and switch to using the source function and radiation field around an optical depth of 0.1.}

An alternative to the above switching scheme is to rewrite the transfer equation in terms of the quantity $K^I=S-I$ (dropping bin indices $i$ here for brevity). The quantity $K^I$ is proportional to the cooling rate in a specific ray direction. The transfer equation then transforms into
\be
\frac{\dd K^I}{\dd \tau} = \frac{\dd S}{\dd \tau}- K^I,
\ee
which in its integral form is given by
\be
K^I(\tau) = K^I(\tau_0) \mathrm{e}^{\tau-\tau_0} + \int_{\tau_0}^\tau \mathrm{e}^{\tau'-\tau} \frac{\dd S}{\dd \tau'}\, \dd\tau'.
\label{eq:Qtransfer-integral}
\ee
The total heating rate is then computed from
\be
Q= \oint  K^I \,\dd\Omega
\ee

This equation does not suffer from the numerical precision problems caused by the cancellation of the subtraction $S$ and $I$. Eq.~\eqref{eq:Qtransfer-integral} has the same form as the formal solution of the normal transfer equation and can be solved efficiently using a variety of methods. If the source function is known (such as when assuming LTE) then solving straight for $K$ is possible without having to solve for $I$ first. \citet{2006A&A...448..731H} describe an elegant method for solving for $K$ in decomposed domains using long characteristics and a direct solution of the transfer equation.  In case of a non-LTE source function, such as in Sect.~\ref{subsec:MGRTS}, then computing $S$ requires solving the transfer equation to obtain $I$ and $J$. Computing $Q$ from $K^I$ afterwards then offers little benefit over computing $Q$ straight from $\divF$ or $S-J$.

\citet{1982A&A...107....1N} implemented a similar method as \citet{2006A&A...448..731H}, but based on the second-order form of the transfer equation:
\be
\frac{\dd^2 P}{\dd \tau^2} = P-S.
\ee
where 
\be
P = \frac{1}{2} \left( I(\Omega) + I(-\Omega) \right),
\ee
 the average of an ingoing and an outgoing ray
 \citep[e.g.,][p. 387]{2014tsa..book.....H}.
Defining $K_P=P-S$ one arrives at
 \be
 \frac{\dd^2 K_P}{\dd \tau^2} = K_P - \frac{\dd^2 S}{\dd \tau^2}.
 \ee
 This equation has the same form as the second order transfer equation and can be solved efficiently along a long characteristic using the method of
 \citet{1964CR....258.3189F}.

\subsection{Summary and examples of photospheric radiative transfer}

\begin{figure}
\centering
 \includegraphics[width=12cm]{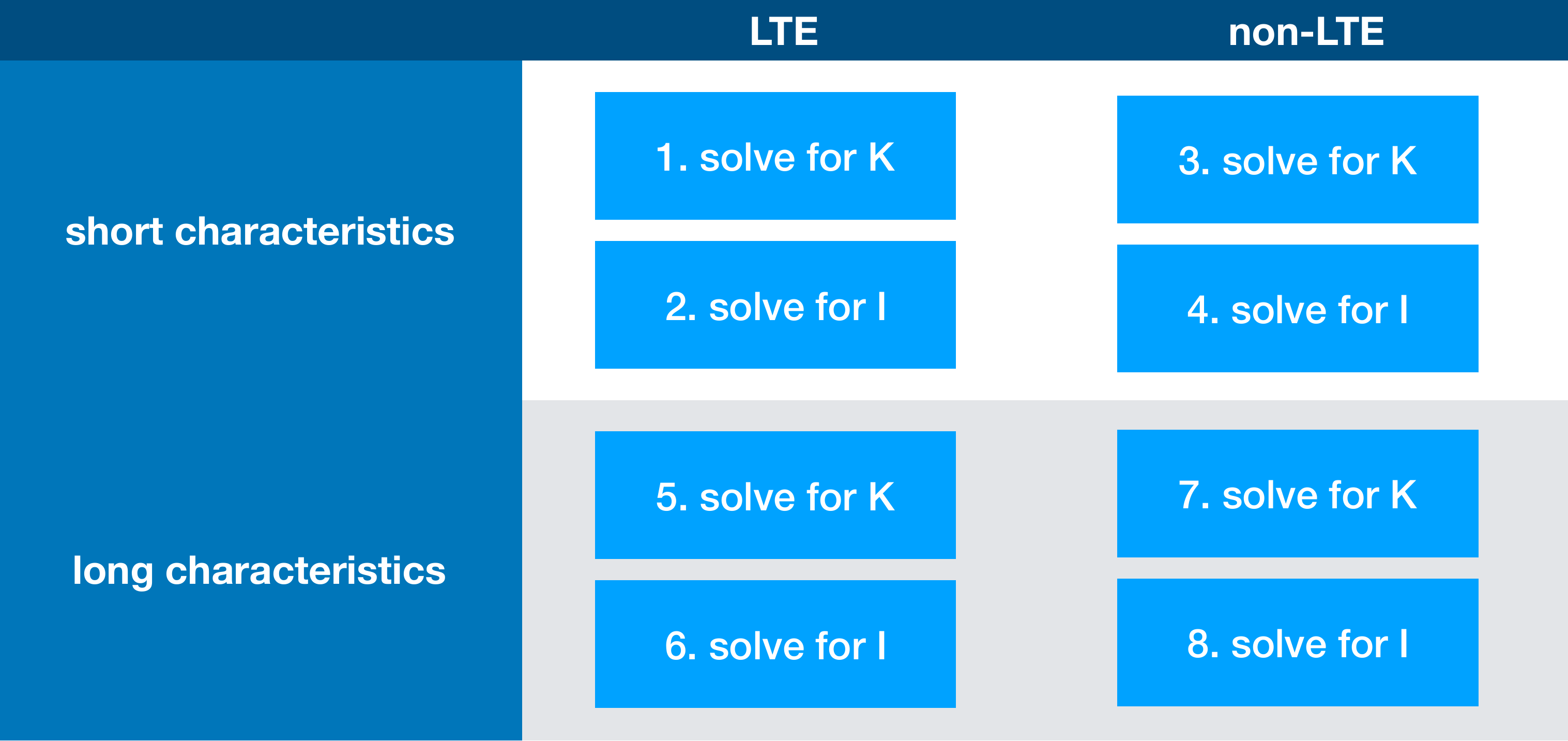}
\caption{Summary table of the three major binary choices in multi-group radiative transfer from computing radiative losses and gains in the photosphere.}
\label{fig:rt-option-table}       
\end{figure}

Figure~\ref{fig:rt-option-table} shows a summary table of the three major binary choices in multi-group radiative transfer from computing radiative losses and gains in the photosphere: LTE or non-LTE source function, short characteristics or long characteristics, and solving for $K$ or solving for $I$ in order to compute the flux divergence. Each of the resulting 8 options is numbered.

The simulation by \citet{1982A&A...107....1N} is of Type 5.  \textsc{MURaM} \citep{2004PhDT-voegler,2017ApJ...834...10R},  \textsc{RAMENS}
\citep{2015ApJ...812L..30I,2016PhDT.........5I}, and \textsc{MANCHA3D}
\citep{2017A&A...604A..66K,2018A&A...618A..87K} are Type 2 codes. \textsc{Stellarbox} \citep{2015arXiv150707999W} is Type 6. COBOLD \citep{2012JCoPh.231..919F} has both Type 2 and Type 6 options, but for stellar atmosphere simulations only Type 2 is supported.

The Oslo version of the \textsc{Stagger} code \footnote{the MHD solver of this code is described in Nordlund \& Galsgaard (1995), see \url{http://www.astro.ku.dk/~kg/Papers/MHD_code.ps.gz}.} (Type 8) and the \textsc{Bifrost} code \citep{2011A&A...531A.154G} (Type 4) are as of October 2019 the only codes using a non-LTE source function. 

\begin{figure}
 \centering \includegraphics[width=\textwidth]{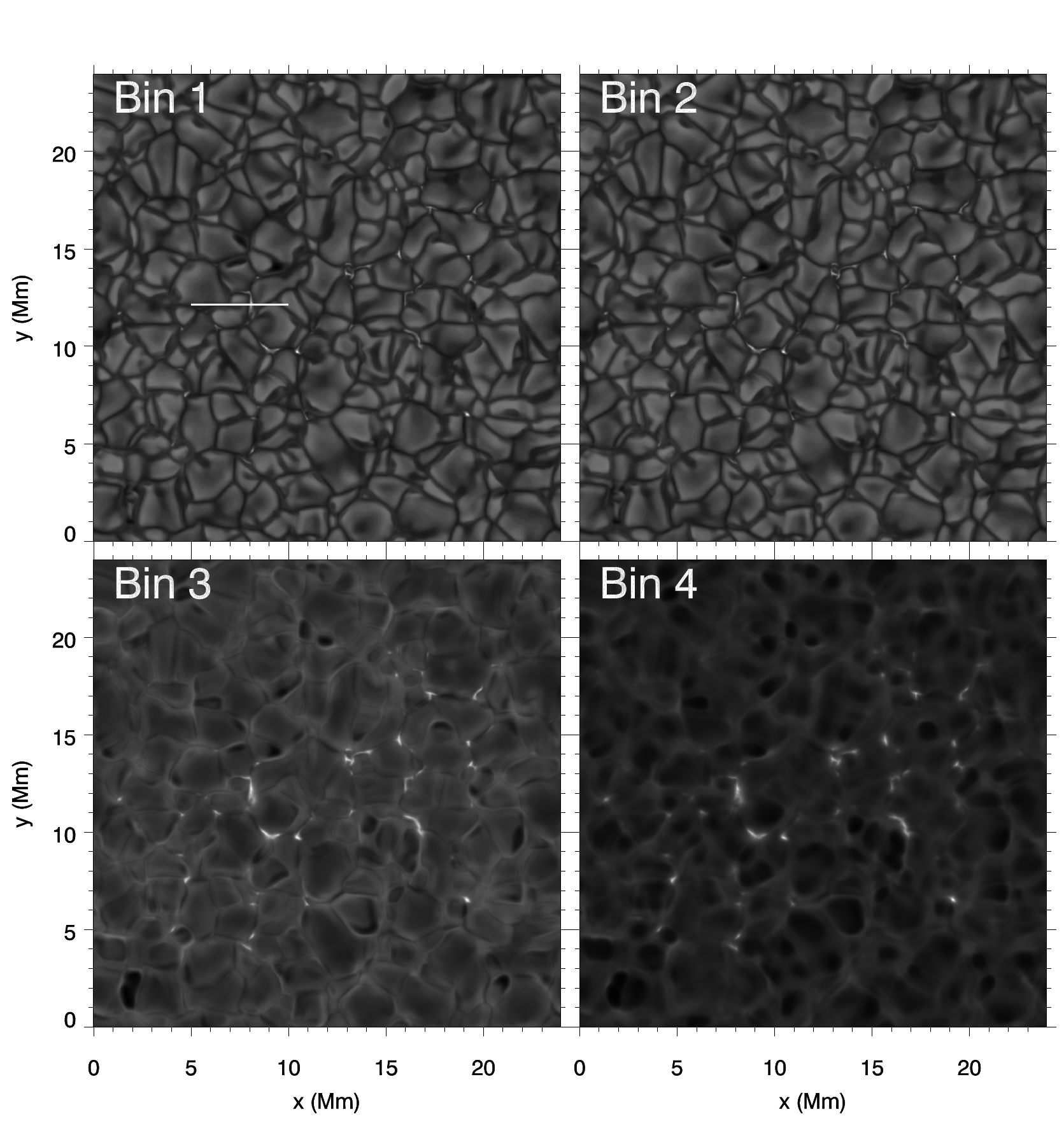}
\caption{Example of the vertically emergent intensity computed with a 4-group non-LTE scheme. The flux divergence along the horizontal white line is shown in Fig.~\ref{fig:bin-heating-per-mass}. The model was computed with the \textsc{Bifrost} code \citep{2011A&A...531A.154G}.}
\label{fig:bin-intensity}       
\end{figure}

\begin{figure}
 \centering
\includegraphics[width=\textwidth]{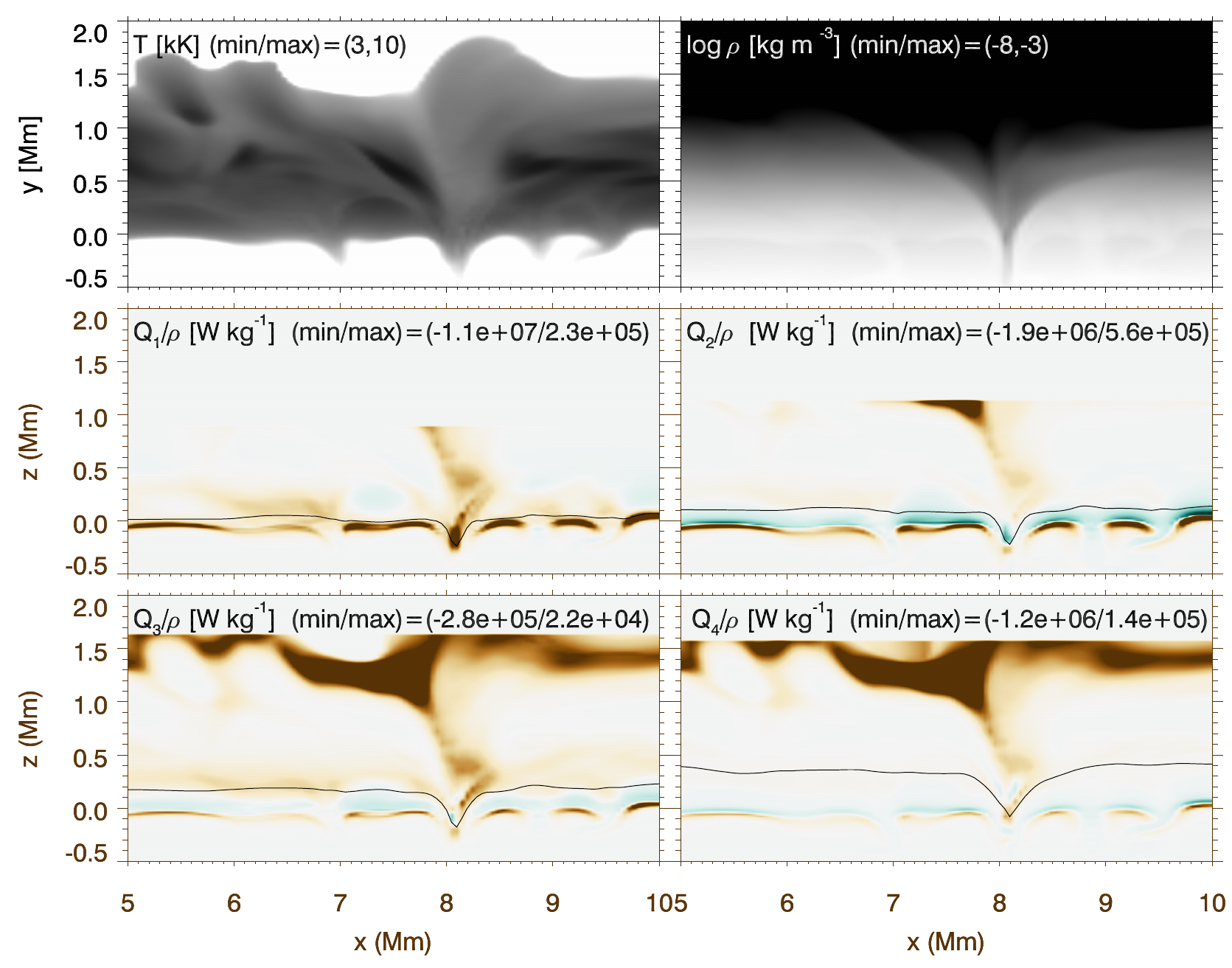}
 \caption{Atmospheric structure and flux divergence per mas unit $Q_i/\rho=-\nabla \cdot \vF_i/\rho = 4 \pi \kappa_i (J_i-S_i)$ in a vertical slice along the white line in Fig.~\ref{fig:bin-intensity}. The top row shows the temperature and density, the two bottom rows show the flux divergence per mass unit in the four opacity groups. Brown is cooling, blue is heating, and the brightness scale for the flux divergence panels is clipped at 20\% of the maximum of the absolute value to enhance contrast.  The maximum and minimum values are indicated in each panel.
 The black curves indicate the $\tau_i=1$ height in each bin. Note that the maxima of the heating and cooling per unit mass do not coincide with the $\tau=1$ height.
The flux divergence in this simulation is artificially set to zero at the height where the entire atmosphere above has a maximum optical thickness of $10^{-5}$. The total radiative losses above this height are negligible, but the local losses and gains per unit mass are not. }
\label{fig:bin-heating-per-mass}       
\end{figure}

Figures~\ref{fig:bin-intensity} and~\ref{fig:bin-heating-per-mass} demonstrate the result of a 4-group computation in a radiation-MHD simulation. The details of this particular simulation can be found in 
\citet{2016A&A...585A...4C}.

Figure~\ref{fig:bin-intensity} shows the vertically emergent intensity in each radiation group. Bins~1 and~2 grouped areas of the spectrum with generally low opacities. The corresponding images are indeed reminiscent of observed optical continuum images of the photosphere. Bins~3 and~4 contain opacities of stronger lines, and resemble images taken in the wings of the \CaII\ H\&K lines
\citep[e.g.,][]{2004A&A...416..333R}.
The images are dominated by reversed granulation; the small bright structures are caused by magnetic field concentrations. 

Figure~\ref{fig:bin-heating-per-mass} shows the heating rate per mass in each radiation bin. The prominent funnel shape in the mass density panel is caused by the presence of a strong magnetic field concentration that fans out with height. All bins show strong cooling at the top of the granules (Red color just below $z=0.0$~Mm), and bins 2\,--\,4 show heating just above the granules. There is strong cooling per mass unit in the chromosphere above $z=1$~Mm. While the optical thickness of the chromosphere above this height is small because of the low density, the heating rate per mass is independent of the mass density, and depends on the value of $\kappa_i$ and the size of $(S_i-J_i)$ only. 

\begin{figure}
\centering
 \includegraphics[width=8.8cm]{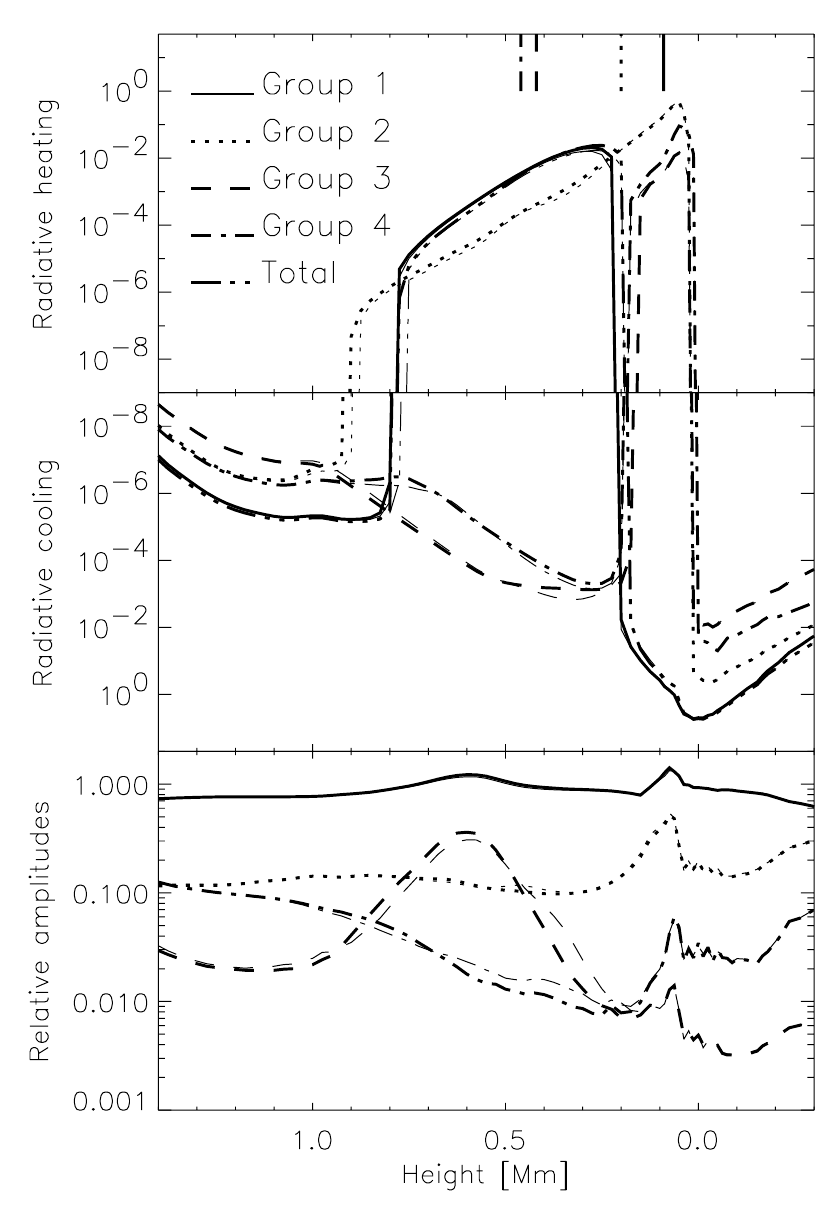}
 \caption{Average radiation heating and cooling per volume ($Q$) as a function of height in a 3D radiation-hydrodynamics simulations using a 4-group non-LTE radiative transfer scheme. Upper panels: Horizontally averaged radiative heating per volume unit shown in a bilogarithmic plot. The units are arbitrary. Thick lines are the exact group solutions, while the thin lines are the approximate group solutions. The vertical line segments in the uppermost panel show the heights where the average optical depth per group is unity, as measured along the vertical line. Lower panel: Horizontally averaged amplitudes of radiative heating (average absolute value) relative to the amplitude of the exact total heating. Thick {curves} are the exact amplitude ratios, while the thin {curves} are ratios from the approximate group solutions.
Adapted with permission from \citet{2000ApJ...536..465S}, copyright by AAS.
\label{fig:group-schem-accuracy}}       
\end{figure}

Figure~\ref{fig:group-schem-accuracy} demonstrates the accuracy of approximating the spectrum by only a few frequency groups in the non-LTE scheme of
\citet{2000ApJ...536..465S}
on the horizontally averaged values of $Q_i$.
It does however not test the assumptions of coherent scattering, LTE equation of state and LTE opacity. The absolute error in $Q = \Sigma Q_i$ is of the order of a few percent, while the error in individual groups $i$ can be as large as 50\% (group 3 at $z=0.3$~Mm). Note that $Q$ is dominated by group 1. The absolute value of $Q$ is decreasing with height because of its dependence on the mass density.

\begin{figure}
\centering
 \includegraphics[width=8.8cm]{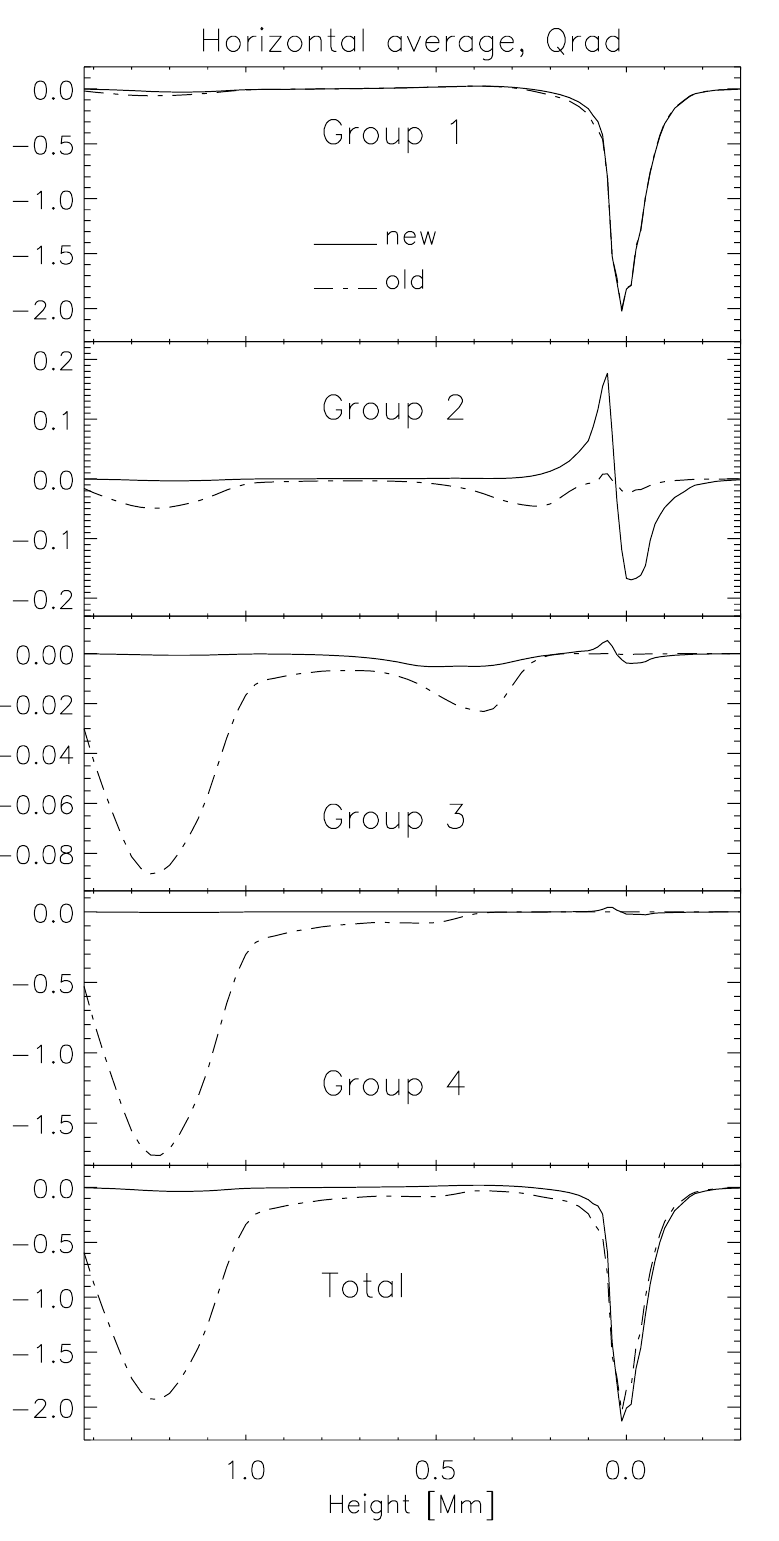}
\caption{Horizontal averages of radiative heating per mass unit ($Q/\rho$ for a 4-group scheme assuming LTE and non-LTE. The units are arbitrary, and the scaling is the same for all figures. The amplitude of the flux divergence for the  LTE method (labeled {\it old}) is much larger above 0.3 Mm than for the non-LTE method (labeleld {\emph new}). This effect is caused mainly by the larger differences between the source function and the mean intensity. Around 0.0~Mm, the new method produces larger amplitudes in the sense that hot regions are cooled more and cooler regions are heated more. Lower panel: Sum of all groups. The main difference above 1.0~Mm comes from the contribution in group 4. Adapted with permission from \citet{2000ApJ...536..465S}, copyright by AAS.
\label{fig:lte-nlte-comp} }     
\end{figure}

In Fig.~\ref{fig:lte-nlte-comp} the difference between assuming an LTE or non-LTE source function is demonstrated. The expression for the flux divergence in the multigroup scheme is the same as for the monochromatic case:
\be
\frac{\divF_i}{\rho} =4 \pi \epsilon_i \kappa_i  \left( S_i - J_i \right),
\ee
where $\epsilon_i = 1$ in LTE, and $\epsilon_i \le 1$ in non-LTE. In the latter case $\epsilon_i$ can be as low as $10^{-4}$. A strict comparison between the scattering and non-scattering cases is not possible, because of the different definition of the group-mean opacities, source functions and thermal emission terms. Nevertheless, the main result from this figure is that the LTE scheme vastly overestimates the radiative cooling in the chromosphere, because $\epsilon_i = 1$ in LTE. 

The amplitude of the cooling and heating in {groups} 2\,--\,4 between $z=0$~Mm and $z=0.3$~Mm is however much larger in non-LTE than in LTE. 
\citet{2000ApJ...536..465S}
speculates that this is caused by the smoothing effect that scattering has on $J_i$, but a thorough investigation of this effect has never been done.

\section{Radiative losses in the transition region and corona} \label{sec:corona}

The corona is optically thin at all wavelengths except for the radio regime. Its radiative losses are dominated by a myriad of EUV lines from highly ionised stages of many different elements
\citep[e.g.,][]{2004A&A...427.1045C,2012SoPh..275..115W}.

In the transition region, which I loosely define here as that part of the atmosphere where hydrogen is ionised but the temperature is below  100~kK, the emission is dominated by lines of lower ionisation stages (typically twice to four times ionised). For most lines, and for most regions on the sun, the TR is optically thin. In solar flares this is not necessarily the case:
\citet{2019ApJ...871...23K}
showed that the TR can have appreciable optical thickness in the \ion{Si}{IV}\ 140~nm lines.

Non-equilibrium ionisation effects play a role in the transition region %
\citep{2013AJ....145...72O,2017A&A...597A.102G}
and corona 
\citep{1993ApJ...402..741H,2004A&A...425..287B,2016A&A...589A..68D}.
Figure~1 of 
\citet{1993ApJ...402..741H}
shows an increase in radiative losses at a given density and temperature in the transition region by a factor two in a 1D hydrodynamic situation when non-equilibrium ionisation is used instead of instantaneous ionisation equilibrium.

A fully general treatment of TR and coronal radiative losses in a radiation-MHD simulation would thus involve solution of the full 3D non-equilibrium-non-LTE radiative transfer problem for most ionisation stages for a wide range of elements. This is currently impossible for 3D simulations because of limits on computation speed.

Neglecting radiative non-local transfer effects through assuming optically thin radiative transfer (i.e., assuming $\Jnu=0$ in the rate equations) alleviates already much of the problems without sacrificing much accuracy in most circumstances. If one furthermore excludes hydrogen and helium, the justification being that these elements are fully ionised at high temperatures and do not contribute to line cooling, then changes in ionisation of the elements do not influence the pressure and temperature.

The problem reduces then to solving the rate equations
\begin{equation}
 \frac{\partial n_i}{\partial t} + \nabla \cdot (n_i\vec{v}) =
 \sum_{j,j \ne i}^{n_{\rml}} n_j P_{ji} - n_i \sum_{j,j \ne i}^{n_{\rml}} P_{ij},
 \label{eq:RateEq2} 
\end{equation}
where $i$ sums over all energy levels and ionisation stages of each element under consideration. The rate coefficients $P_{ij}$ contain collisional (de-)excitation and collisional ionisation recombination terms and spontaneous radiative deexcitation and recombination terms, but no radiative terms that involve absorption of an existing photon.
Ignoring absorption therefore only allows for cooling. The radiative loss rate per volume in a bound-bound transition between a lower level $i$ and upper level $j$ is then given by
\be
Q_{ij} = h \nu_{ij} A_{ji} n_j,
\ee
and a similar expression can be written for bound-free transitions. The total cooling rate per volume can then be computed by summing the contributions of all transitions of all elements and including electron-ion free-free radiation. A more detailed description of this method in a 1D radiation-hydrodynamics code is given in
\citet{1993ApJ...402..741H}.
To my knowledge this method has not been implemented in 3D codes. 

The option of computing non-equilibrium ionisation was added to the \textsc{Bifrost} code by
\citet{2013AJ....145...72O},
but they did not implement the resulting radiative cooling.

Instead, the default method to compute radiative losses in the corona is to assume statistical equilibrium (i.e., the left-hand-side of Eq.~\eqref{eq:RateEq} is assumed to be zero. Together with the assumption of no photon absorption these two assumptions together are often called the ``coronal approximation''. The cooling a a spectral line of element $X$ in ionisation stage $m$ with upper level $j$ and lower level $i$ can then be written as:
\bea
Q_{ij}&=& h \nu_{ij} \frac{A_{ji}}{\nel} \, \frac{n_{j,m}}{n_m} \, \frac{n_m}{n_X} \, \frac{n_X}{\nH} \, \nel \nH, \\
 & \equiv & G(T,\nel) \, \nel \nH.
\eea
Here $n_{j,m}/n_m$ is the fraction of all ions in ionisation stage $m$ in level $j$, $n_m/n_X$ is the fraction of all atoms of species $X$ in ionisation stage $m$, and $n_X/\nH$ is the abundance of element $X$ relative to hydrogen. The function $G(T,\nel)$ is only weakly dependent on the electron density: the upper level population is dominated by collisional excitation from the ground state, so that $\nel n_m \sim A_{ji} n_{j,m}$, and the rate coefficients setting up the ionisation balance are almost linear in the electron density.  A residual electron density dependence remains in $G(T,\nel)$ through collisional deexcitation and density-dependent dielectronic recombination 
\citep{1972MNRAS.158..255S,1974MNRAS.169..663S}.
%

\begin{figure}
\centering
 \includegraphics[width=0.8\textwidth]{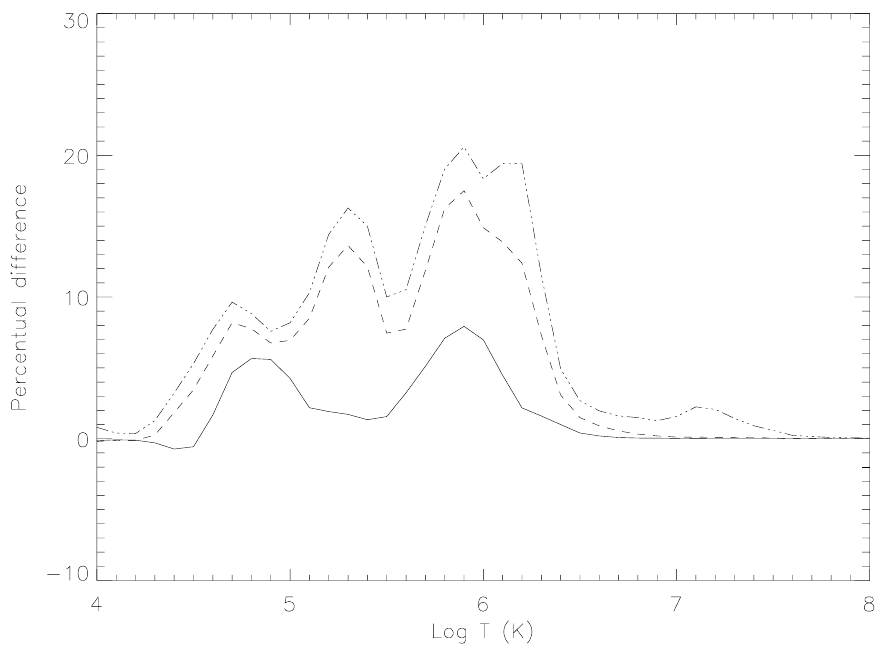}
\caption{Sensitivity of the coronal loss function $\Lambda(T,\nel)$ to the electron density. Full line: $10^{14}$ vs. $10^{16}$~m$^{-3}$; dashed line: $10^{14}$ vs. $10^{18}$~m$^{-3}$; dash-dotted line: $10^{14}$ vs. $10^{20}$~m$^{-3}$. Adapted with permission from \citet{1999A&A...347..401L}, copyright by ESO.}
\label{fig:lambda-ne-sensitivity}       
\end{figure}

A total coronal radiative loss function $\Lambda(T,\nel)$ can thus be constructed through summing up $Q_{ij}$ for all levels and ionisation stages of all relevant elements and adding the contribution from continuum processes. The electron density dependence of $\Lambda(T,\nel)$ is weak: 
\citet{1999A&A...347..401L}
performed a sensitivity study for electron densities in the range $10^{14}$\,--\,$10^{20}$~m$^{-3}$, and found a difference {in the value of  $\Lambda$} of at most 20\% (see Fig~\ref{fig:lambda-ne-sensitivity}). It is therefore reasonably accurate to pre-compute $\Lambda$ at a fixed typical coronal electron density (say $\nel=10^{16}$~m$^{-3}$).

A larger uncertainty can be introduced by inaccuracies of the elemental abundances. The coronal loss function is dominated by losses from C, Si, O, and Fe, and the abundances of these elements have a large influence. The abundance of C and O is debated after 3D non-LTE {computations} 
\citep{2004A&A...417..751A,2005A&A...431..693A}
gave a different result than calculations using 1D models
\citet{1998SSRv...85..161G}.
%

\begin{figure}
\centering
 \includegraphics[width=0.8\textwidth]{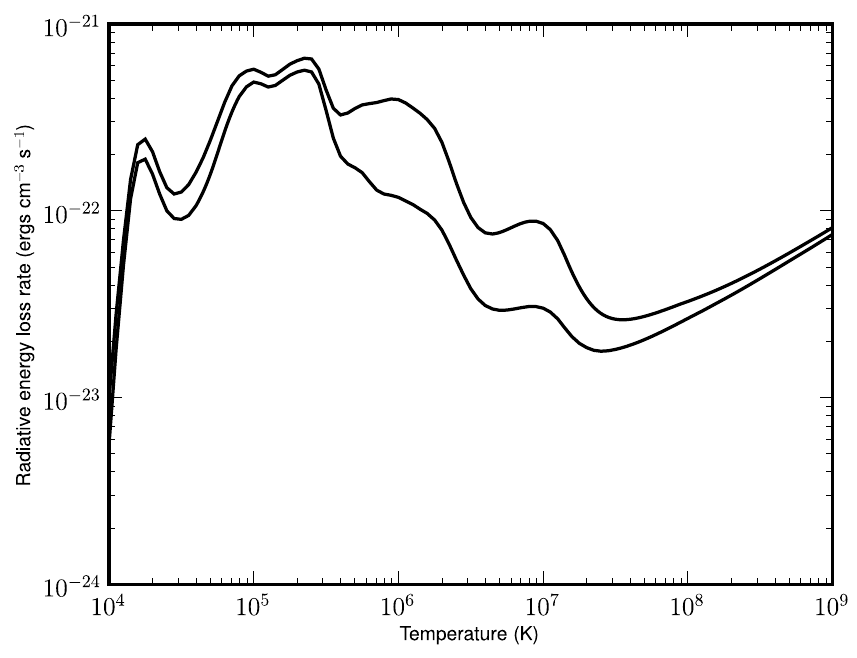}
\caption{Radiative loss functions $\Lambda$ computed assuming $\nel=10^{15}$~m$^{-3}$ as function of temperature. Upper curve: coronal abundances from \citet{1992ApJS...81..387F}. Lower curve: photospheric abundances from  \citet{1998SSRv...85..161G}. Adapted with permission from \citet{2009A&A...498..915D}, copyright by ESO.}
\label{fig:radiative-loss-function}       
\end{figure}

Another complication is that the most accurate abundances are derived from photospheric lines, but coronal abundances are generally different from abundances in the photosphere.
\citet{1992ApJS...81..387F} compared coronal abundances for 15 elements to photospheric abundances,
and Fig.~\ref{fig:radiative-loss-function} show the difference in the loss function when using photospheric or coronal abundances. Unfortunately it appears that no systematic re-investigation of coronal abundances have been made since 1992. Furthermore, the coronal abundances are not constant, but depend on the coronal structure 
\citep[see for example Sec 2.1 of ][and references therein]{2019ARA&A..57..157C}.
Abundances might thus well be the dominant source of uncertainty in the coronal loss function. 

The CHIANTI atomic database\footnote{\url{http://www.chiantidatabase.org}} 
\citep{1997A&AS..125..149D,2019ApJS..241...22D}
provides an extensive compilation of critically assessed atomic data. In addition to the data it delivers software packages in both the IDL and Python languages for using this data and easy generation of loss functions using a variety of abundances and other input options.

Computation of $Q$ in the corona is then straightforward:
\be
Q = - \Lambda(T) \nel \nH, \label{eq:coronal-loss-function}
\ee
where $\Lambda(T)$ is computed using a 1D lookup table.
The hydrogen density is easily computed from the mass density, and the electron density can be approximated accurately assuming full ionisation of both H and He at high temperatures. In the low temperature regime, where \HI, \HeI, and \HeII\ exist in significant amounts, one can employ precomputed tables of the electron density assuming LTE or coronal equilibrium.

In order to avoid contribution from the convection zone, photosphere, and chromosphere, one needs to multiply $Q$ with a cutoff function that drives Q to zero in the lower atmosphere. \textsc{Bifrost} employs a soft cutoff function: $\mathrm{exp}(-P/P_0)$, with $P$ the gas pressure and $P_0$ a typical pressure at the top of the chromosphere
\citep{2011A&A...531A.154G}.
\textsc{MURaM} uses a hard cutoff at T=20,000~K
\citep{2017ApJ...834...10R}.

The radiative losses in the TR and corona tend to have sharp peak in the transition region owing to the quadratic density-dependence. 
\citet{2017ApJ...834...10R}
noticed that the relatively large grid spacing of the simulations compared to the thickness of the TR can lead to inaccuracies in the computation of $Q$. He proposed a scheme using subgrid interpolation to improve the accuracy.

\begin{figure}
\centering
 \includegraphics[width=\textwidth]{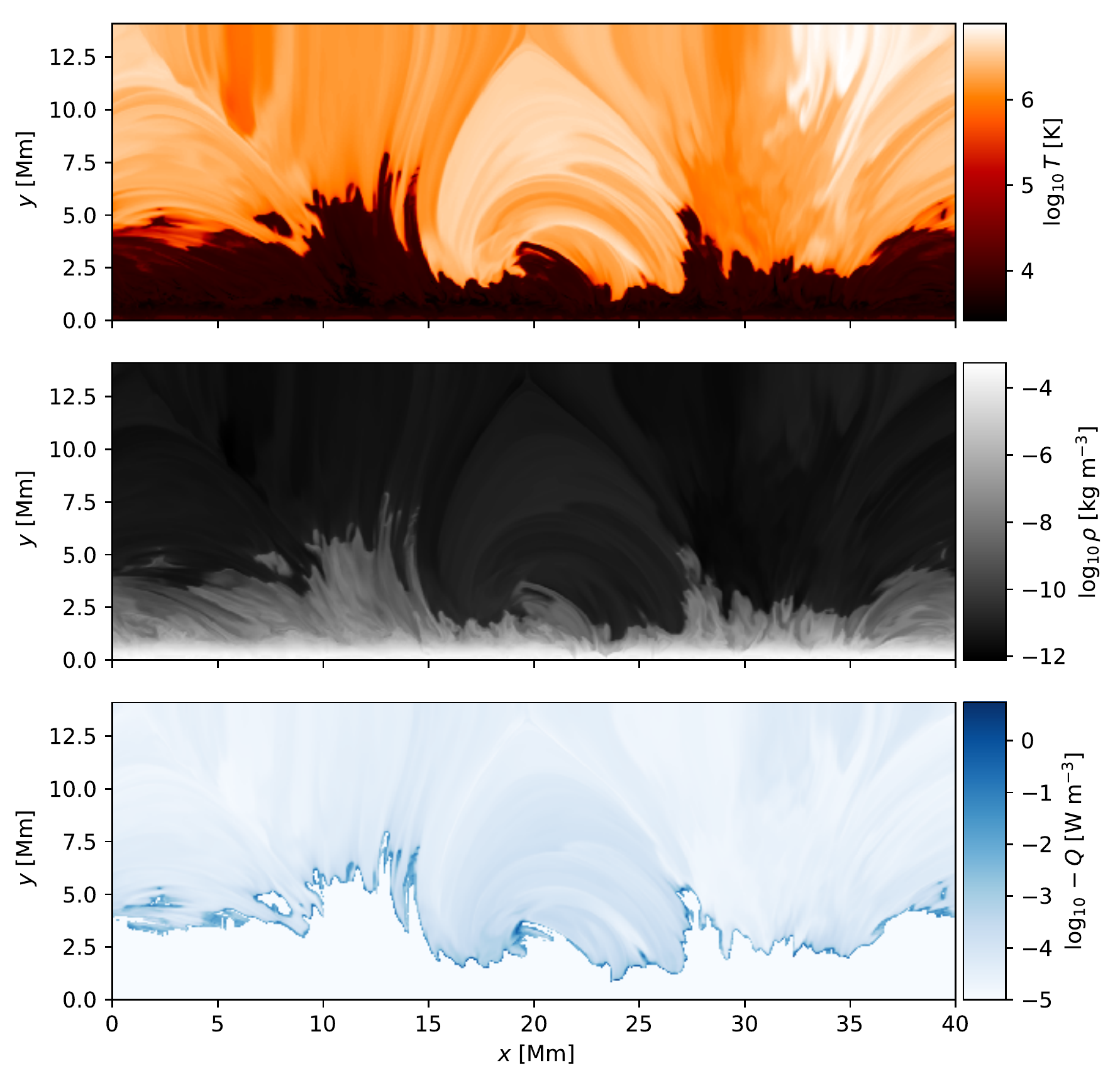}
\caption{Radiative losses in the transition region and corona in a radiation-MHD simulation computed with the \textsc{MURaM} code. Top: temperature; middle: mass density; bottom: $-Q$, i.e., a positive value means radiative cooling. Also note the strong corrugation of the chromosphere-TR boundary. Figure computed from a simulation by Danilovic et al. (2019).}
\label{fig:qloss-MURaM}       
\end{figure}

A demonstration of $Q$ and its relation to temperature and density is shown {in} Figure~\ref{fig:qloss-MURaM} for a simulation carried out with the \textsc{MURaM} code. The cooling is largest (note the logarithmic scale) in the transition region, just at the border between the TR and the chromosphere, where it typically is in the range of 0.1-1.0~W~m$^{-3}$, and it quickly drops down to coronal values around $10^{-4}$~--~$10^{-5}$~W~m$^{-3}$. Zooming in reveals that the strongly-emitting layer is often only a few pixels wide.

\section{Radiative transfer in the chromosphere} \label{sec:chromosphere}

\begin{figure}
\centering
 \includegraphics[width=\textwidth]{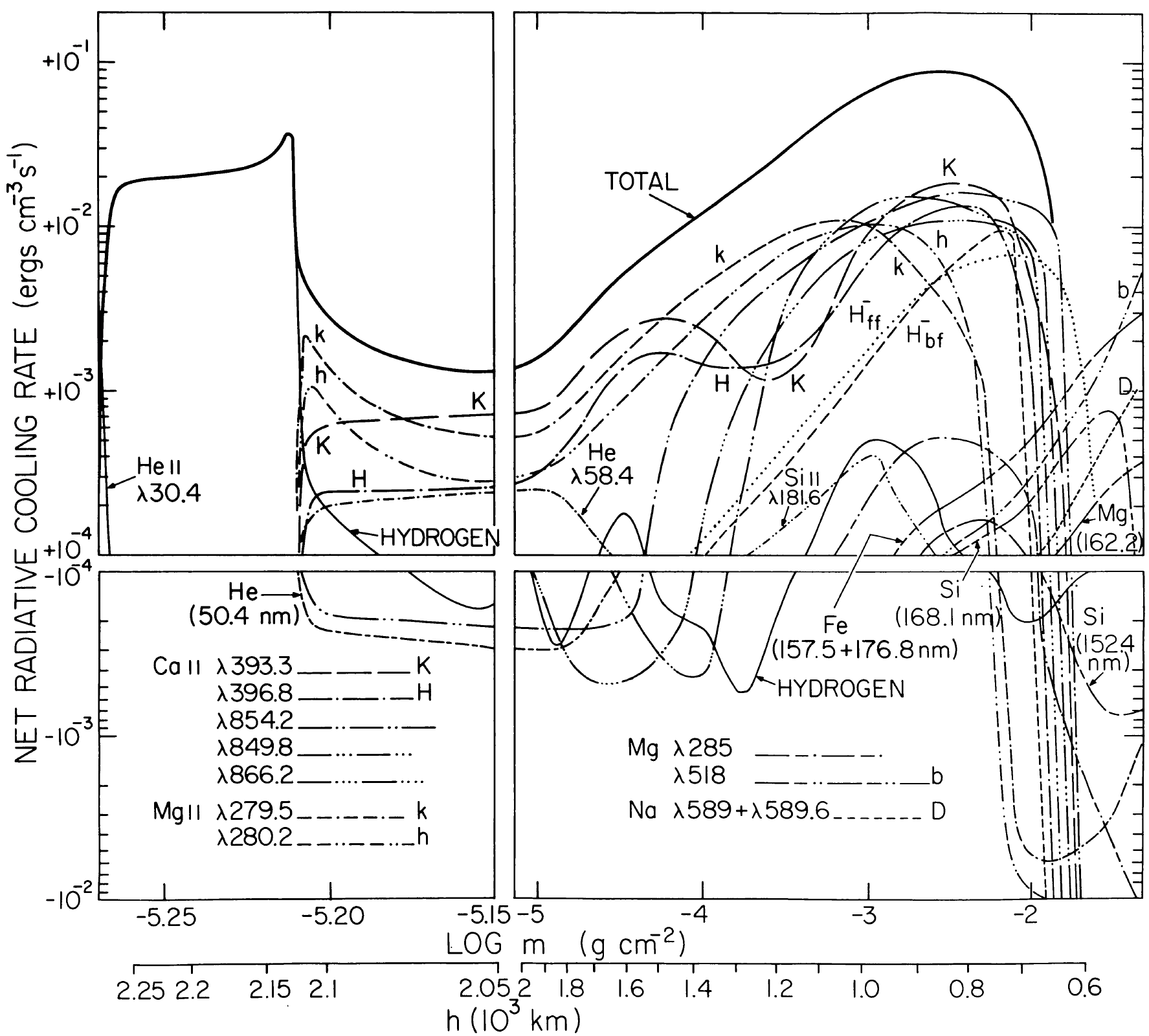}
\caption{Net radiative cooling in the 1D semi-empirical VAL3C model atmosphere. The cooling between $z=700$~km and $z=2120$~km in this model is dominated by five lines from \CaII\ and two lines from \MgII. At larger heights \HI\ Ly$\alpha$ alone is the dominant radiative cooling agent. Adapted with permission TODO:ASK PERMISSION from \citet{1981ApJS...45..635V}, copyright by AAS.
\label{fig:falc-heating} }     
\end{figure}

The assumptions underpinning the methods for photospheric radiative transfer as presented in Sect.~\ref{sec:multigroup} are no longer valid in the chromosphere: the chromosphere has a low opacity except in a few strong spectral lines, { in ultraviolet continua
        below 160~nm, and in (sub-)mm continua above 160~$\mu$m.} The lines that  dominate the radiative energy exchange are the H\&K and infrared triplet of \CaII, the h\&k lines of \MgII, and \HI\ Ly$\alpha$ (see Fig.~\ref{fig:falc-heating}). The opacity in these lines is severely underestimated in the construction of a bin-averaged opacity. The source function is no longer described by the Planck function, and the assumptions of coherent scattering {or complete redistribution are} very inaccurate for these strong lines, which should instead be modelled using partially coherent scattering (PRD). The ionisation balance of hydrogen and helium is out of equilbrium.

Proper inclusion of the radiative cooling in the chromosphere (even when ignoring non-equilibrium effects) thus involves solving the 3D non-LTE radiative statistical-equilbrium transfer problem including PRD. This is in principle possible using dedicated radiative transfer codes, but it is computationally expensive. A single solution to the problem for a single atom costs a CPU time of  $\sim10$~s per grid cell
\citep{2017A&A...597A..46S},
which is very large compared to the $\sim5$~$\mu$s CPU time per grid cell per timestep for radiation-MHD simulations
\citep[e.g.,][]{2011A&A...531A.154G}.
Simplifications that speed up the computation are thus required. 

\citet{2012A&A...539A..39C}
developed techniques to do so by describing the net effect of all the radiative transfer as a combination of (1) an optically thin radiative loss function which represents the local energy loss through radiation per atom in the right ionisation stage per electron, (2) the probability that this energy escapes the atmosphere, and (3) the fraction of atoms in the ionisation stage under consideration. These three factors must all be determined empirically because there are no obvious general physics-based approximations.

The method approximates the radiative loss per volume owing to species $X$ in ionisation state $m$ as
\be
Q_{X_m} = - L_{X_m} E_{X_m}(\tau) \frac{n_{X_m}}{n_X} A_X \frac{n_\mathrm{H}}{\rho} n_\mathrm{e} \rho.
\ee
Here, $L_{X_m}$ is the optically thin radiative loss function per electron and per particle of element $X$ in ionisation stage $m$, $E_{X_m}(\tau) $ is the photon escape probability as function of some depth parameter $\tau$, $\frac{n_{X_m}}{n_X}$ is the fraction of element $X$ in ionisation stage $m$, and $A_X$ the abundance. 

The quantities $L_{X_m}$, $E_{X_m}$, and $n_{X_m}/n_X$ are determined from a 1D  radiation-hydrodynamic simulation including non-equilibrium ionisation computed with the Radyn code
\citep{1992ApJ...397L..59C,2002ApJ...572..626C} for \HI. For \CaII\ and \MgII\ they were determined from a 2D radiation-MHD simulation with \textsc{Bifrost} that provided the atmospheric structure and subsequent statistical equilibrium radiative transfer calculations including PRD using Multi3d 
\citep{2009ASPC..415...87L}. 
%

\begin{figure}
\centering
 \includegraphics[width=0.8\textwidth]{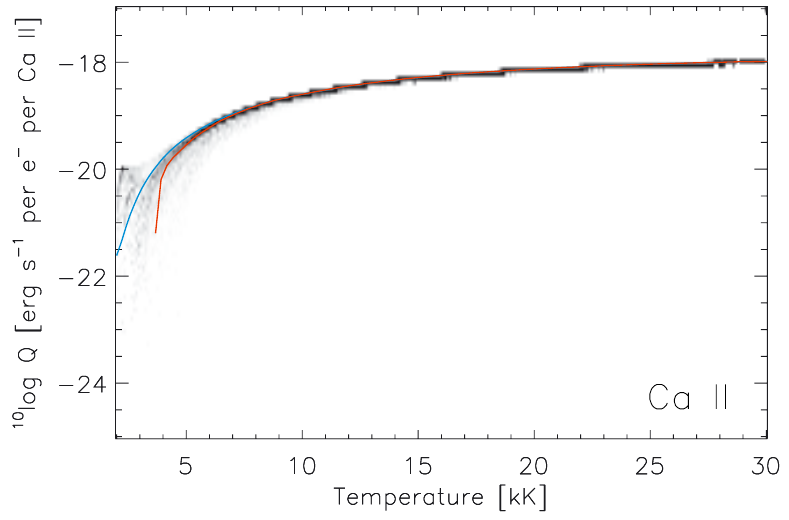}
  \includegraphics[width=0.8\textwidth]{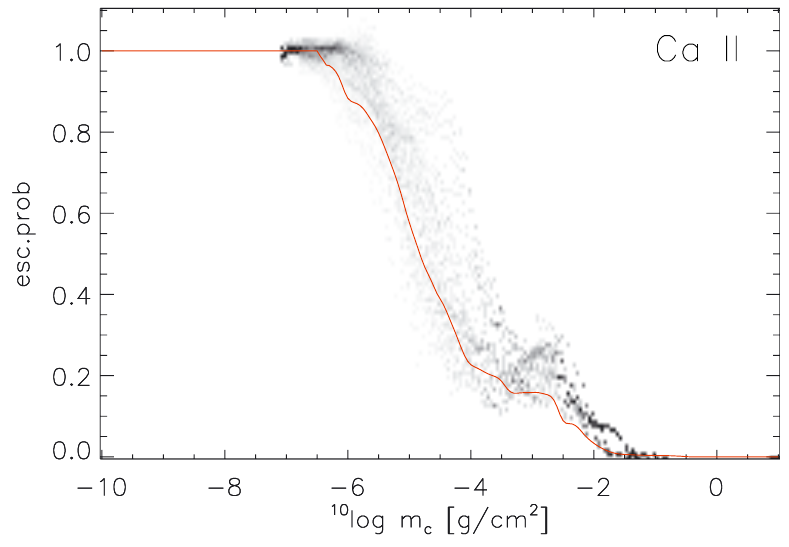}
   \includegraphics[width=0.8\textwidth]{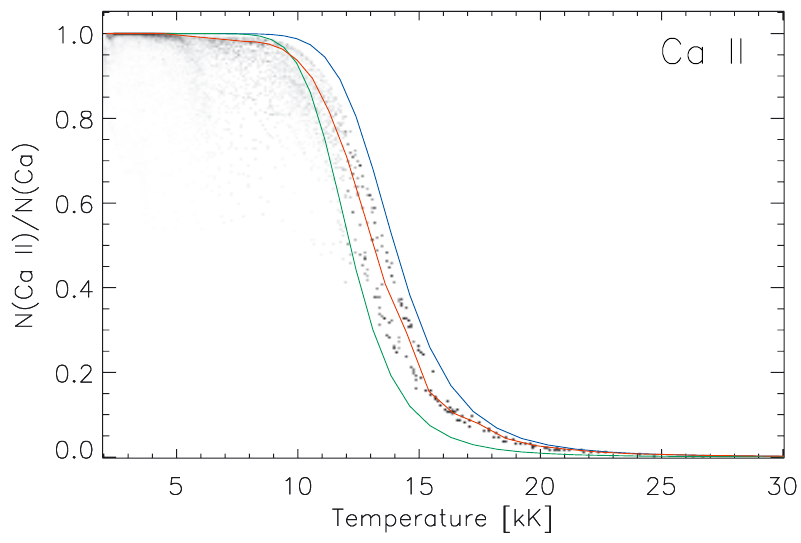}
\caption{Chromospheric radiative losses in \CaII\ using empirically calibrated recipes. Top: JPDF of radiative loss function and temperature in the  chromosphere of a radiation-MHD model. The blue curve is the coronal approximation, the red curve the adopted fit. \CaII\ actually heats (meaning a negative loss function, not shown here because of the logarithmic axis) at low temperatures, and that is why the red curve appears a bad fit. Middle: escape probability, with the adopted fit in red. Bottom: fraction of all Ca as \CaII. Red is the adopted fit, blue the ionisation balance under coronal equilibrium conditions and green the balance assuming LTE.  Adapted with permission from \citet{2012A&A...539A..39C}.}
\label{fig:CL2012ingredients}     
\end{figure}

The loss function  $L_{X_m}$ can be computed for each grid cell in the simulation by summing the net downward radiative rates multiplied with the energy difference of the transition, summed over all relevant transitions. The joint probability density function (JPDF) of $L_{X_m}$ and gas temperature for \CaII\ is shown in the top panel of Fig.~\ref{fig:CL2012ingredients}. Radiative transfer effects make that $L_{X_m}$ is no longer a unique function of $T$, but the red curve indicates an approximate fit. The ionisation degree can be computed from the atomic level populations in the calibrating simulations (see the bottom panel of Fig.~\ref{fig:CL2012ingredients}). Again they are no longer clean functions of temperature, but the spread is minor and a sensible fit as function of temperature can be made. Finally, the empirical escape probability $E_{X_m}$ is computed from the total radiative losses in the simulation, the radiative loss function and the ionisation degree. The middle panel of Fig.~\ref{fig:CL2012ingredients} shows $E_{\CaII}$ with the vertical column mass as depth parameter. Again there is considerable spread in the JPDF.

\begin{figure}
\centering
 \includegraphics[width=8.8cm]{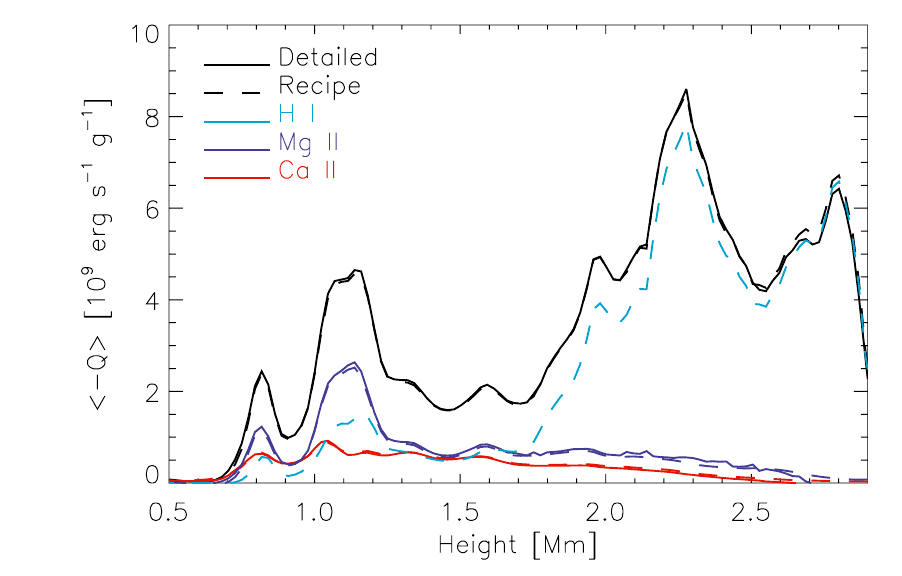}
\caption{Test of the validity of tabuled chromospheric radiative losses. The curves show the horizontally averaged radiative cooling (i.e., $-Q/\rho$ as a function of height in a 2D \textsc{Bifrost} radiation-MHD simulation. The colored lines show the results computed from a 2D non-LTE radiative transfer computation (solid) and the approximate recipe (dashed) are shown. For hydrogen there is no detailed radiative transfer computation because the recipe was computed from a 1D simulation with the Radyn code. The black solid land dashed lines show the total cooling from all three elements combined. Adapted with permission from \citet{2012A&A...539A..39C}.}
\label{fig:chromo-recipe-test}     
\end{figure}

Figure~\ref{fig:chromo-recipe-test} displays a comparison of this simple recipe with a detailed calculation. The recipe does a surprisingly good job in reproducing the average cooling given the simplicity of the method. However, in individual grid cells large errors in $Q$ might occur, caused by the spread of values around the red curves in Fig.~\ref{fig:CL2012ingredients}.

Computing this chromospheric radiative loss function in a radiation-MHD simulation is fast and simple; it only involves computing the values of the fitted quantities from 1D lookup tables. The electron density can for example be determined from a 2D lookup table computed assuming LTE. This is not particularly accurate but simple. A more realistic electron density can be obtained by computing the non-equilibrium ionisation balance of hydrogen and/or helium (see Sect.~\ref{sec:noneq-ionisation}).

\citet{2012A&A...539A..39C}
also describe techniques to model the absorption by the chromosphere of radiation emitted in the corona. Half of this radiation is emitted downwards and the majority will be absorbed in the chromosphere. The corona emits mainly in the far UV regime, and the dominant source of extinction at these wavelengths are the \HI, \HeI, and \HeII\ continua. The coronal loss function is a frequency-integrated quantity (see Sect.~\ref{sec:corona}), so in order to model the extinction one has to estimate a representative opacity. \citet{2012A&A...539A..39C} propose to use the opacity at the edge of the \HeI\ continuum at 50~nm. The contribution to the flux divergence is then given by
\be
Q_\mathrm{abs} = 4\pi \kappa_{\mathrm{He I}} \rho J_\mathrm{cor},
\ee 
with $\kappa_{\mathrm{He I}}$ the representative opacity. The angle-averaged radiation field $J_\mathrm{cor}$ is computed from the transfer equation:
\be
\frac{\dd I}{\dd s} = \eta - \kappa_{\mathrm{He I}} \rho I,
\ee
with the emissivity given by $\eta = - Q_\mathrm{cor}/4\pi$. The quantity $Q_\mathrm{cor}$ is the coronal loss function as defined in Eq.~\eqref{eq:coronal-loss-function}. In other words: it is assumed that only the corona contributes to the production of photons, the chromosphere can only absorb these photons, and scattering is ignored. 

Because most radiation-MHD codes already include a 3D radiative transfer module used for the photospheric radiation, one can relatively easily implement the absorption of coronal radiation at a very modest increase in computation time. Simpler 1D recipes that ignore the spreading of radiation in the horizontal directions are also possible. They offer little else besides a marginal increase in computation speed, and can lead to spuriously large heating when coronal radiation emitted by a small localized source is forced to be absorbed in a single vertical column.

The population of neutral helium in the ground state needed to compute $\kappa_{\mathrm{He I}}$ can be computed assuming LTE populations without a too large error. Computing it taking into account non-equilibrium ionisation is more accurate but much slower (see Sect.~\ref{sec:noneq-ionisation}).

\begin{figure}
\centering
 \includegraphics[width=\textwidth]{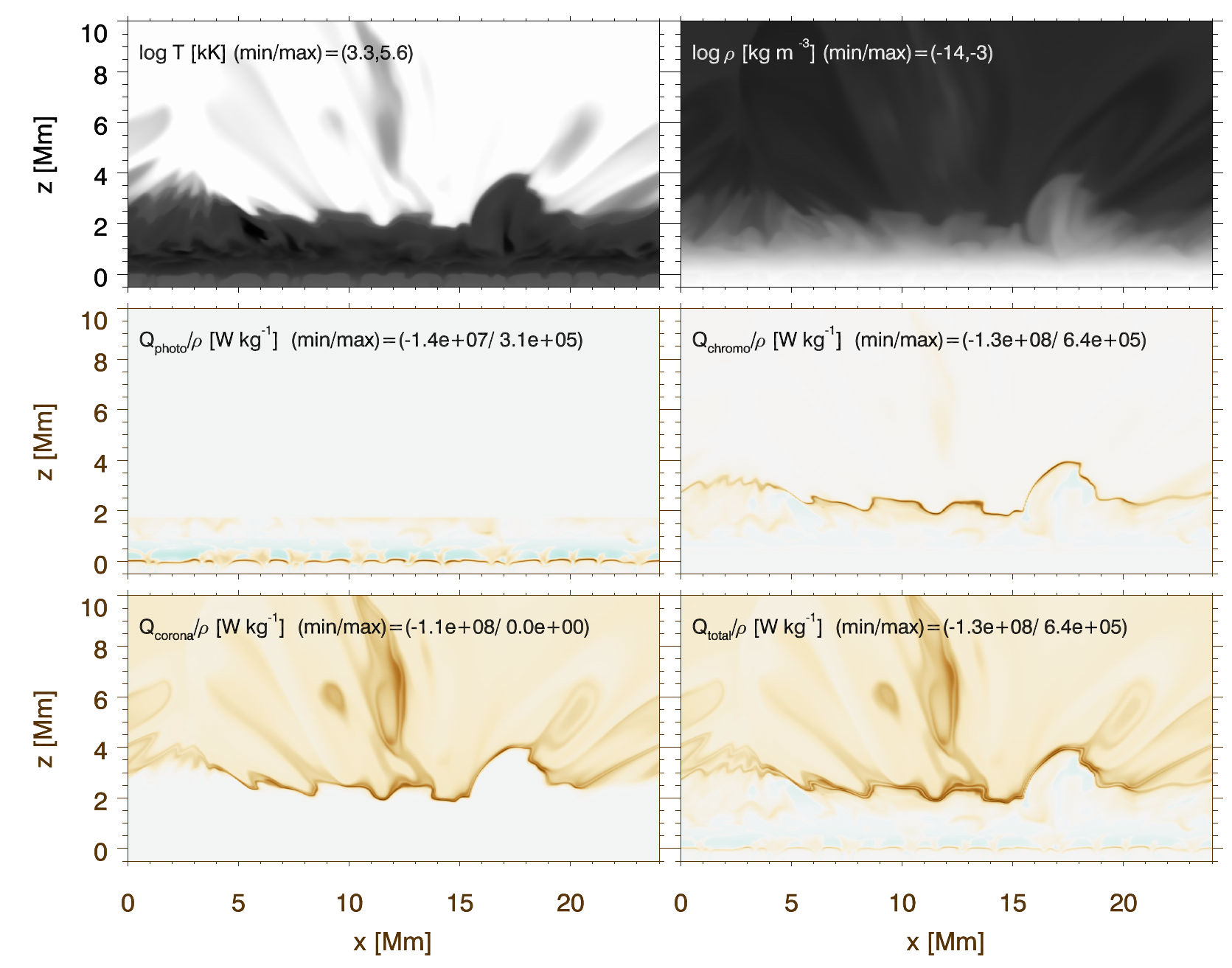}
\caption{Example of chromospheric radiative losses and gains in comparison to the photospheric and coronal losses and gains in a 3D simulation computed with the \textsc{Bifrost} code. Top row: temperature and density in a vertical slice through the simulation. The panel labeled $Q_\mathrm{photo}$ shows the radiative losses {per mass unit} computed with the method described in Sect.~\ref{sec:multigroup}; {$Q_\mathrm{chromo}$} displays the chromospheric losses and absorption of coronal radiation {per mass unit}  using the methods from \citet{2012A&A...539A..39C};  $Q_\mathrm{corona}$ shows the losses {per mass unit}  computed as described in Sect.~\ref{sec:corona}. The panel labeled $Q_\mathrm{total}$ shows the sum of the previous three panels, i.e., all radiative losses and gains in the simulation. {Brown color indicates cooling, blue color represents heating.}}
\label{fig:chromoQ}     
\end{figure}

Figure~\ref{fig:chromoQ} gives an example of how the various approximations for radiative cooling act in different atmospheric regimes. The photospheric cooling rate per mass is largest in the photosphere, but has a small contribution up into the chromosphere. Coronal losses are relevant throughout the entire corona, but are largest in the transition region. The chromospheric losses are largest just below the transition region owing to Ly$\alpha$ cooling (see also Fig.~\ref{fig:falc-heating}). Modest cooling owing to \CaII\ and \MgII\ lines and heating from the absorption of coronal radiation happens somewhat deeper in the chromosphere. The lower-right panel, which combines photospheric, chromospheric, and coronal losses, clearly shows that the largest radiative losses per mass unit occur in the transition region. 


\section{The equation of state and non-equilibrium ionisation}
\label{sec:noneq-ionisation}

\begin{figure}
\centering
 \includegraphics[width=0.7\textwidth]{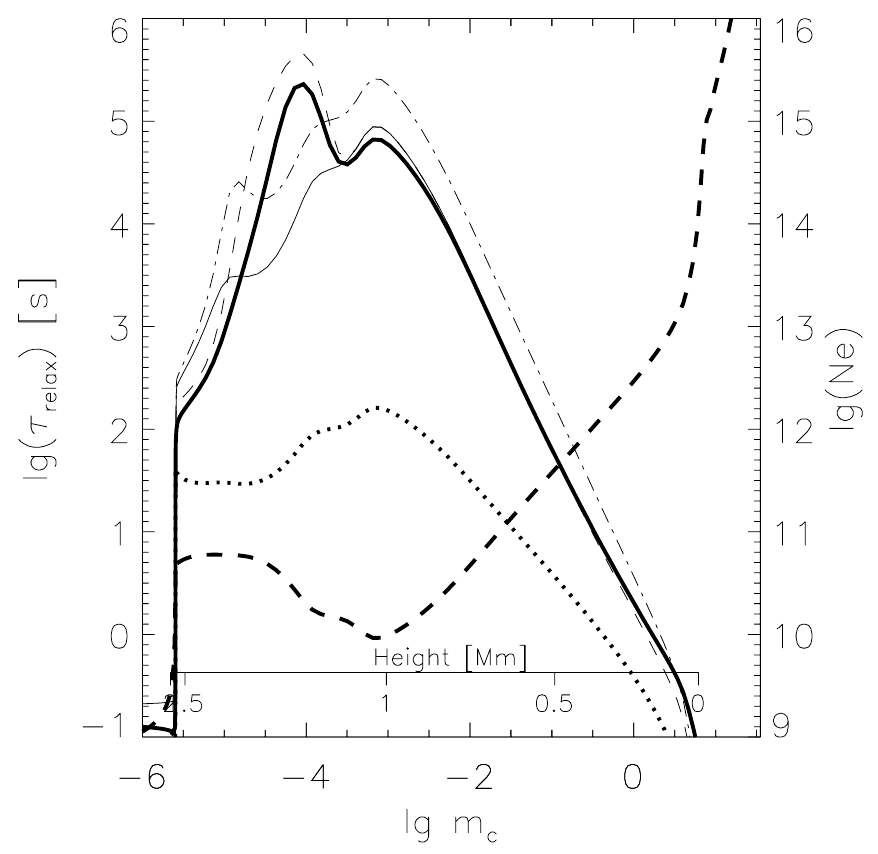}
\caption{Relaxation timescale for hydrogen ionisation as function of column mass in a 1D radiation hydrodynamics simulation computed with the RADYN code. The numerically determined relaxation timescale is given as the thick solid line. Timescales determined from eigenvalues of the rate matrix are also shown for several cases: full rate matrix (dotted line), collisions only (dot-dashed line), Lyman transitions in detailed balance (thin solid line), and Ly$\alpha$ treated with a constant net radiative bracket (dashed line). The electron density is given as the thick dashed curve. Adapted with permission from \citet{2002ApJ...572..626C}, copyright by AAS.}
\label{fig:cs2002}     
\end{figure}

Non-equilibrium radiative transfer and non-equilibrium ionisation play a role in the solar atmosphere wherever magneto-hydrodynamic timescales are short compared to timescales on which an atomic system reacts to changes. 
The combined continuity and rate equation for the level population $n_i$ in level $i$ of an atomic species with $N$ energy levels is given by: 
\be \frac{\partial n_i}{\partial t} + \nabla \cdot (n_i\vec{v}) =
 \sum_{j=1,j \ne i}^N n_j P_{ji} - n_i \sum_{j,j \ne i}^N P_{ij},
 \label{eq:RateEq}
\ee
where $P_{ij}$ is the rate coefficient for transitions from level $i$ to level $j$. If one ignores the advection term and assumes that all the $P_{ij}$ are constant {in time} then the equations for all levels together form a set of coupled first-order differential equations
\be
\frac{\partial \vec{n}}{\partial t} = P\vec{n}, 
\ee
with $\vec{n}$ the vector of level populations and $P$ the rate coefficient matrix. This system has the solution:
\be
\vec{n}(t) = \sum_{i=1}^N c_i \vec{a}_i \rme^{\lambda_i t},
\ee
where $a_i$ are the eigenvectors of the rate matrix with corresponding eigenvalues $\lambda_i$, and $c_i$ are constants that depend on the initial condition. One of the eigenvalues is zero, and the corresponding eigenvector represents the equilibrium solution. All other eigenvalues are negative and have an absolute value smaller than one. If the system starts away from the equilibrium solution, then it will evolve towards equilibrium on a characteristic timescale given by
\be
\tau = \frac{1}{\min(|\lambda_i|)}, \lambda_i \ne 0.
\ee
A detailed discussion of the time dependence of atomic level populations is given in 
\citet{2005JQSRT..92..479J}.
One of the results from that paper is that the slowest time scale tend to be associated with processes that have small net rate coefficients. The smallest rate coefficients are often associated with ionisation and recombination processes (as well as transitions involving metastable levels).

An illustrative solution for a two-level atom was derived in
\citet{2002ApJ...572..626C}.
The equilibration timescale is given by
\be
\tau = 1 \left/ \left( C_{21}+C_{12} + R_{21} \left[ 1 - \frac{n_1}{n_2} \frac{R_{12}}{R_{21}}\right] \right) \right..
\ee
with $C_{12}$ and $C_{21}$ the upward and downward collisional rate coefficient, $R_{12}$ and $R_{21}$ the radiative rate coefficients and the term between square brackets the net radiative bracket. The collisional coefficients depend linearly (or quadratically in case of collisional recombination) on the electron density and exponentially on temperature. For hydrogen {one} thus expects a maximum in the chromosphere where the temperature and density are relatively low and strong lines are close to detailed balance so that the net radiative bracket is small.

\citet{2002ApJ...572..626C}
did a detailed calculation of this timescale for hydrogen and found that the timescale can be as long as $10^5$~s in the chromosphere (see Fig.~\ref{fig:cs2002}), and that this is the timescale on which ionisation equilibrium is established.

A similar calculation for helium was done for helium by 
\citet{2014ApJ...784...30G},
finding timescales of the order $10^2$~s to $10^3$~s in the chromosphere and transition region, again associated with the ionisation equilibrium

These timescale are larger than the hydrodynamical timescales in the chromosphere and transition region. This has severe consequences for the treatment of radiation hydrodynamics. Because hydrogen and helium are majority species they do not only contribute to the radiative flux divergence, but their ionisation state also influences the temperature, pressure and electron density. The  assumption that the equation of state (EOS) can be computed assuming LTE is no longer accurate. 

The relaxation timescale itself depends on the radiation field, so that a proper treatment of the EOS now requires solution of Eq.~\eqref{eq:RateEq} together with an equation for charge conservation
\be
 n_\mathrm{e} = \sum_{i,j,k} (j-1) \nijk \label{eq:chargeeq},
\ee
and energy conservation
\be
e = \frac{3}{2} \kb T \left(\sum_{i,j,k} \nijk + \nne \right) + \sum_{i,j,k}  \nijk E_{ijk}  \label{eq:energyeq}.
\ee
Here $\nijk$ are the atomic level populations of species $i$ in ionisation state $j$ and excitation state $k$ and $E_{ijk}$ is the sum of the dissociation, ionization and excitation energy of a particle in state $ijk$. The rate coefficients $P_{ij}$ contain the radiation field so that the transport equation (Eq.~\eqref{eq:transfer_form1}) must be solved too. As stated before, this is too computationally expensive to be of use in 3D simulations.

\citet{sollum1999}
developed approximations to the chromospheric radiation field in hydrogen transitions based on detailed 1D calculations with the Radyn code. He found that the angle-averaged and profile-function-averaged radiation field in a given transition can be approximated by a constant above a certain height in the chromosphere, and by the local Planck function at larger depth. In between these two limits he specifies a smooth transition. His recipes allow for an extreme simplification because the solution of the coupled set of Eqs.~\eqref{eq:RateEq},~\eqref{eq:chargeeq}, and~\eqref{eq:energyeq} can then be solved without having to solve the transfer equation. A limitation of his method is that the Lyman-$\alpha$ line was set in detailed {radiative} balance. This limitation makes the solution less accurate in the very upper chromosphere and transition region.

\begin{figure}
\centering
 \includegraphics[width=\textwidth]{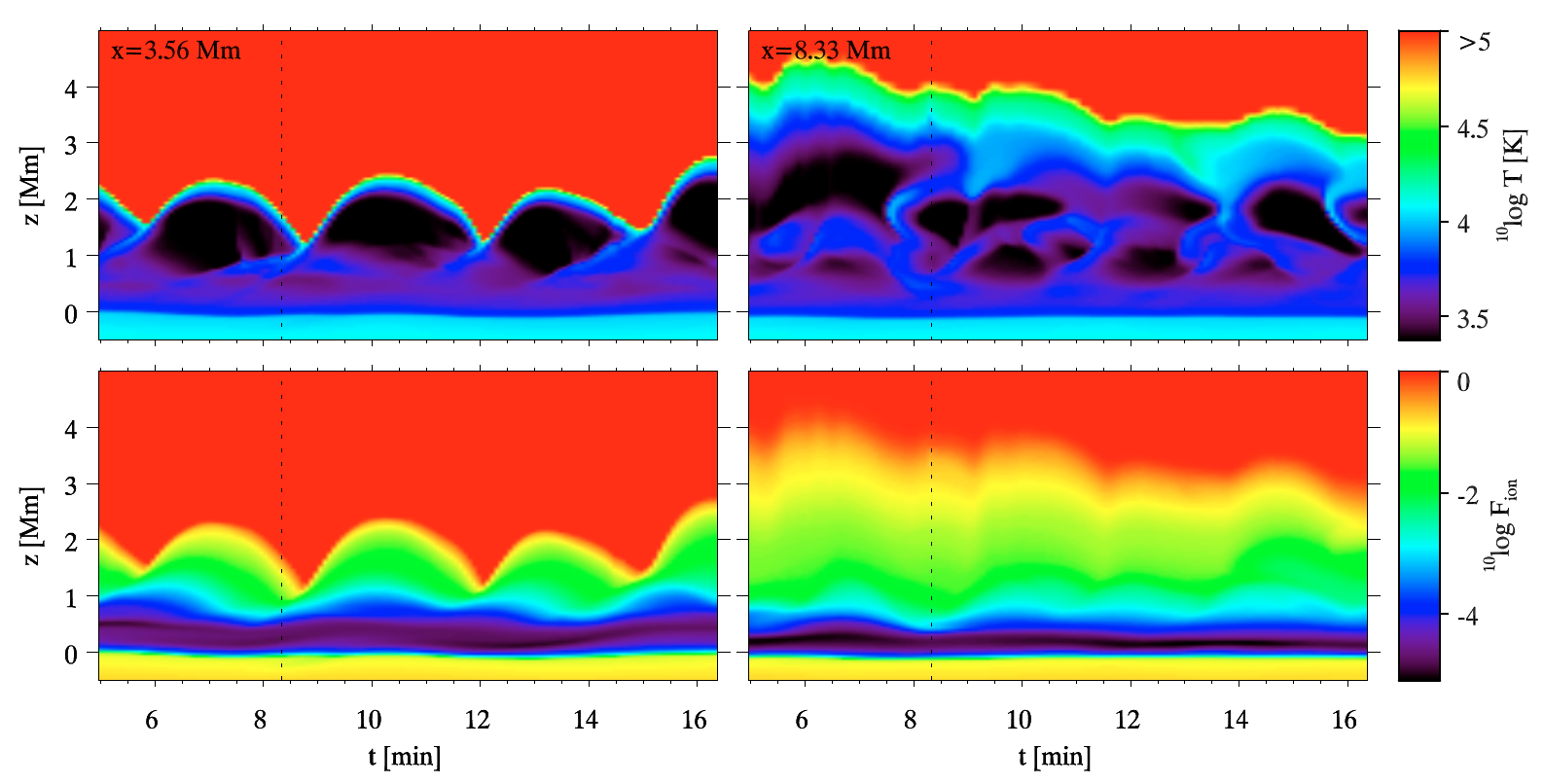}
\caption{Temperature and hydrogen ionisation degree $n_\mathrm{\HI} /(n_\mathrm{\HI} + n_\mathrm{\HII})$ as function of time in two columns of a 2D radiation-MHD simulation that includes non-equilibrium ionisation of hydrogen. Left-hand panel: a column in a magnetic element. Right-hand panel: a column with weak magnetic field. The left-hand panel shows regular shock waves with 3-minute period, while the right-hand panel shows a more irregular temperature structure. In both cases the ionisation degree does not follow the temperature structure. Instead it is rather constant.
 Adapted with permission from \citet{2007A&A...473..625L}, copyright by ESO.}
\label{fig:leenaarts++2007}     
\end{figure}

His method was implemented in the Oslo \textsc{Stagger} Code by 
\citet{2007A&A...473..625L},
and used to perform a 2D radiation-MHD simulation of the solar atmosphere with a non-equilibrium EOS. Helium was still treated  assuming LTE populations. The temperature and hydrogen ionisation degree in this simulation is demonstrated in Fig.~\ref{fig:leenaarts++2007}. As a consequence of the long equilibration timescale, the ionisation degree in the chromosphere does not follow the temperature. The ionisation degree is instead rather constant in time for a given Lagrangian fluid element.

\citet{2016ApJ...817..125G} 
developed approximations for the radiation field in helium transitions. The physics of helium ionisation is somewhat more complicated than for hydrogen because ionisation in the chromosphere is {mainly} driven by  UV radiation produced in the corona. The resulting recipes take this into account, but require the solution of the transfer equation in seven radiation bins in order to approximate the radiation field in the continua of helium as well as the \HeII~30.4~nm line. 

These authors also implemented a semblance of radiative transfer in the Ly~$\alpha$ line. They assigned a single-bin source function based on the net radiative rate and representative opacity to the line, and used those to solve the transfer equation ignoring scattering. The resulting radiation field is then used to add an upward radiative rate coefficient in Eq.~\eqref{eq:RateEq}. In this way the recipes by 
\citet{sollum1999}
were extended to the very upper chromosphere, but in a rather crude fashion.

The increased realism of an EOS that includes non-equilibrium ionisation comes at a rather steep price in computational efficiency: A simulation with a non-equilibrium EOS is around three (hydrogen only) to five times (hydrogen and helium) slower than a similar simulation with an LTE EOS.

Non-equilibrium ionisation has strong consequences on the structure of the chromosphere and transition region. Most importantly, ionisation can no longer function as efficiently as an energy buffer when the internal energy density of a gas parcel changes. 

\begin{figure}
\centering
 \includegraphics[width=0.8\textwidth]{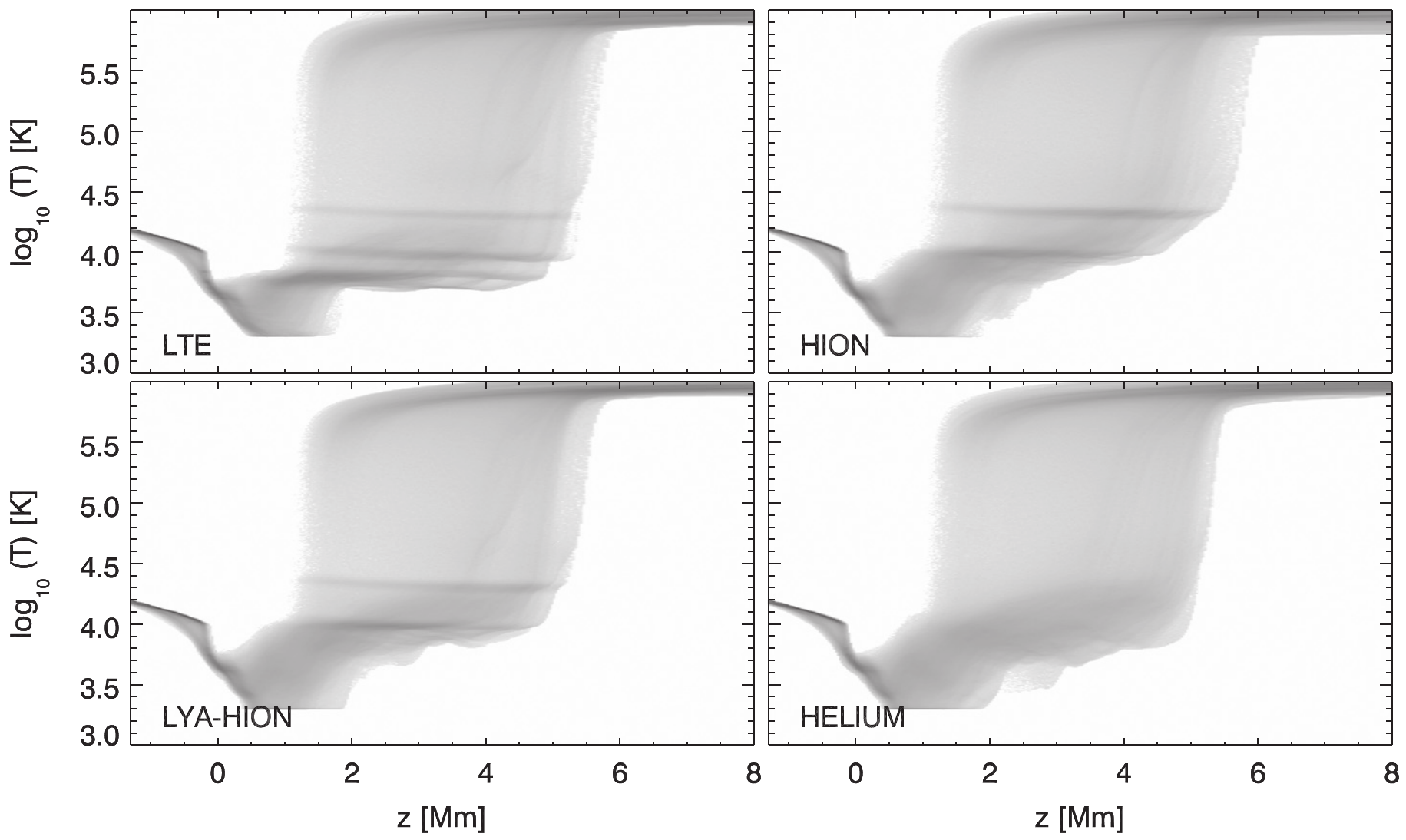}
\caption{Joint probability density functions of height and temperature in four radiation-MHD simulations of the solar atmosphere. The simulations differ in their treatment of the equation of state. LTE: LTE equation of state; HION: hydrogen in non-equilibrium, helium in LTE, hydrogen Lyman transitions in detailed balance.  LYA-HION: as HION but with radiative transfer in Lyman transitions; HELIUM: both hydrogen and helium in non-equilibrium. The three horizontal plateaus (at 6, 10, and 22 kK) in the LTE simulation indicate preferred temperatures when using the LTE equation of state. These temperatures are associated with the LTE ionisation of H I, He I, and He II. The plateaus vanish when we introduce non-equilibrium hydrogen and helium ionisation. Adapted with permission from \citet{2016ApJ...817..125G}.}
\label{fig:golding2016}     
\end{figure}

In LTE energy must be expended to ionise hydrogen and helium {if} the internal energy density is increased before the temperature can rise. Likewise, a decrease in internal energy leads to an instantaneous transfer of ionisation energy to thermal energy, slowing down the temperature decrease. This leads to characteristic bands of ''preferred temperatures'' in joint probability density functions of radiation-MHD simulations assuming LTE (see Fig.~\ref{fig:golding2016}). The temperature {of} these bands are associated with the temperatures where \HI, \HeI, and \HeII\ ionise according to the Saha--Boltzmann equations.

If the ionisation balance is computed in non-equilibrium, and the ionisation/recombination timescale is long, then an increase in internal energy leads directly to a temperature increase. Temperature decreases are also stronger than in LTE because ionisation energy cannot be released quickly enough to counteract cooling. The bands of preferred temperature disappear, and the gas in the chromosphere behaves somewhat like and ideal gas. A clear example of this effect is the increased the amplitude of the temperature jump in acoustic shocks
\citep[see ][]{2002ApJ...572..626C,2007A&A...473..625L}.

Low-temperature areas in the chromosphere have a higher electron density than predicted by the Saha--Boltzmann equations, and the reverse is true for high-temperature areas. This has en effect on the formation of chromospheric lines because the source function couples to the local temperature mainly through collisions with electrons. Non-equilibrium ionisation is also expected to have an effect on the efficiency of heating through ambipolar diffusion
\citep[e.g.,][]{2017Sci...356.1269M,2018A&A...618A..87K}.
%

\section{Other developments}
\label{sec:new-developments}

\subsection{Fast approxmiate radiative transfer in the photosphere}
\begin{figure}
\centering
 \includegraphics[width=0.5\textwidth]{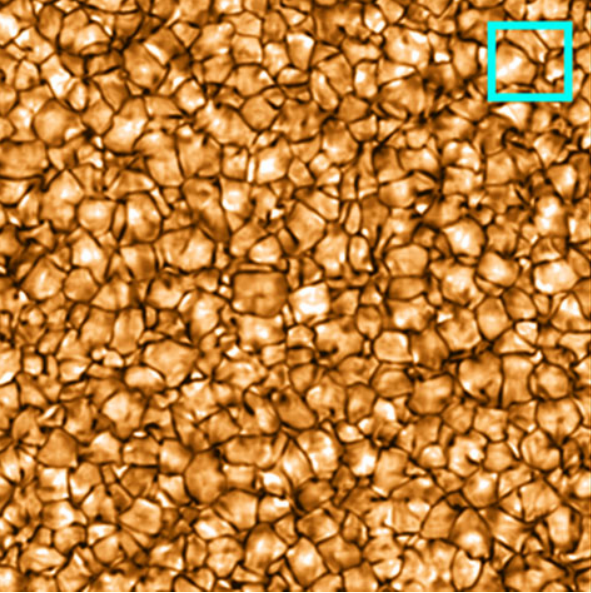}
\caption{Temperature at a constant height in the photosphere of a 3D simulation computed with the RADMHD code. The panel spans a size of $24 \times 24$~Mm$^2$ and has a grid spacing of 21.3~km. The convection pattern strongly resembles the result of simulations that model radiation with higher fidelity, despite the strong simplifications in the computation of the radiative losses. Adapted with permission from \citet{2012SoPh..277....3A}, copyright by Springer.}
\label{fig:abbett2012}     
\end{figure}

\citet{2012SoPh..277....3A} propose a method for quick evaluation of the radiative losses in the photosphere as an alternative to the methods described in Sect.~\ref{sec:multigroup}. They assume that each column in a simulation can be treated as an independent plane-parallel atmosphere. The angle-averaged radiation field at vertical optical depth $\tau_\nu$ in a column is then given by:
\be
\Jnu(\tnu) = \frac{1}{2} \int_0^\infty \Snu(\tnu') E_1 \left(|\tnu-\tnu' | \right) \dd\tnu'.
\ee
The first exponential integral $E_1(x)$ peaks sharply at $x=0$, so that this expression can be approximated as
\bea
\Jnu(\tnu) & \approx & \frac{1}{2} \Snu(\tnu)  \int_0^\infty E_1 \left(|\tnu-\tnu' | \right) \dd\tnu' \\ 
&\approx&	 \Snu(\tnu)  \left( 1- \frac{E_2(\tnu)}{2} \right)
\eea
This result is inserted into the equation for the radiative cooling (Eq.~\eqref{eq:novelfluxdiv}) to yield
\be
Q \approx - 2 \pi \rho \int_0^\infty  \kappa_\nu \Snu E_2(\tnu) \, \dnu.
\ee
If one assumes an LTE source function, then the frequency integral can be approximated using a reasoning similar to the one given in Sect.\ref{subsec:MGRTL}, and together with some additional assumptions one arrives at the final result:
\be
Q \approx - 2 \kappa^B \rho \sigma T^4 E_2(\tau^B),
\ee
where $\kappa^B$ is the Planck-averaged opacity (see Sect.~\ref{subsec:MGRTL}),  $\tau^B$ the optical depth computed from this opacity, and $\sigma$ is the Stefan--Boltzmann constant. With this scheme, computation of the radiative heating only requires a simple 2D table lookup to get $\kappa^B$, and a column-by-column integration over depth to compute $\tau^B$. 

Figure~\ref{fig:abbett2012} shows the resulting temperature structure in the photosphere in a simulation where the radiative losses are computed in this simplified fashion. The method is simple and extremely fast and is well suited for problems that do not require very accurate radiative losses, but nevertheless want the radiative losses to drive reasonably realistic-looking convection. 

\subsection{Escape probability method} \label{sec:escape-prop}

An interesting new method that speeds up non-equilibrium radiative transfer was proposed by 
\citet{2017ApJ...851....5J}.
The major time consuming task in radiative transfer is the evaluation of the formal solution for all required frequencies and angles in order to construct the angle-averaged and frequency-averaged radiation field 
\be
\bar{J} = \frac{1}{4 \pi} \int_0^\infty \int_\Omega \Inu  \, \dd\Omega \, \dnu, \label{eq:Jbar}
\ee
which appears in the radiative rate coefficients of Eq.~\eqref{eq:RateEq}.

\citet{2017ApJ...851....5J}
suggests to compute $\bar{J}$ from the equation
\be
\frac{\dd}{\dd \tau} \left(S-\bar{J}\right) = q^{1/2} \frac{\dd}{\dd \tau} \left( q^{1/2} S \right), \label{eq:judge2017}
\ee
\citep[see][]{1975MNRAS.173..167F,1982ApJ...263..925H}, 
where $\tau$ is a vertical optical depth parameter and the function $q$ an escape probability function that only depends on the vertical optical depth. This equation is approximate only: the main approximations are that the source function varies only slowly over an optical path length and that horizontal structure in the atmosphere can be ignored. 

The big advantage of using Eq.~\eqref{eq:judge2017} is that it replaces the repeated evaluations of the transfer equation in order to compute $\bar{J}$ in a transition with the solution of a single integral along vertical columns in the atmosphere. Solving the statistical equilibrium non-LTE radiative transfer problem problem in a \textsc{MURaM} test atmosphere is $\sim100$ times faster than the full method. 

\citet{2017ApJ...851....5J}
shows that this method can also be used to solve non-equilibrium problems (i.e, it can be used to simultaneously solve Eqs.~\eqref{eq:RateEq}, \eqref{eq:chargeeq}, and~\eqref{eq:energyeq}). Radiation-MHD simulations using this method to compute chromospheric radiative energy exchange or the equation state have so far not been reported on. It would be very interesting to see whether the method is fast enough to be used in practice and how it compares to the methods described in Sect.~\ref{sec:noneq-ionisation}.

\section{Conclusions and outlook}

In this review I presented the most commonly used methods to approximate the transfer of energy between solar plasma and the radiation field in radiation-MHD simulations. The theory and methods for computing radiative energy exchange in the photosphere are perhaps the most well-developed and well-studied. They are also the most accurate:  \citet{2013A&A...554A.118P} compared a 3D hydrodynamics model of the upper solar convection zone that employed accurate abundances and opacity calculations, together with multi-group LTE radiative transfer with 11 groups. The resulting model reproduces the observed continuum center-to-limb intensity variation, the absolute flux spectrum, the wings of  hydrogen lines and the distribution of continuum intensities caused by granulation to a high degree of precision. It seems therefore safe to say that we can model photospheric radiative energy exchange sufficiently accurate for almost all purposes.

Radiative losses in the transition region and corona are much less accurate when using the standard method of an optically loss curve computed from statistical equilibrium, no absorption of radiation and a fixed set of abundances as described in Sect.~\ref{sec:corona}. Inaccuracies caused by ignoring non-equilibrium ionisation effects appear to be largest in the transition region and during solar flares, and might well lead to a factor two error in the value of the loss function $\Lambda(T)$. The choice of abundance has a smaller effect in the transition region, but can lead to differences up to a factor three at higher temperatures in the corona. A systematic test of these effects in 3D radiation-MHD simulations has so far not been done.

\begin{figure}
\centering
 \includegraphics[width=12cm]{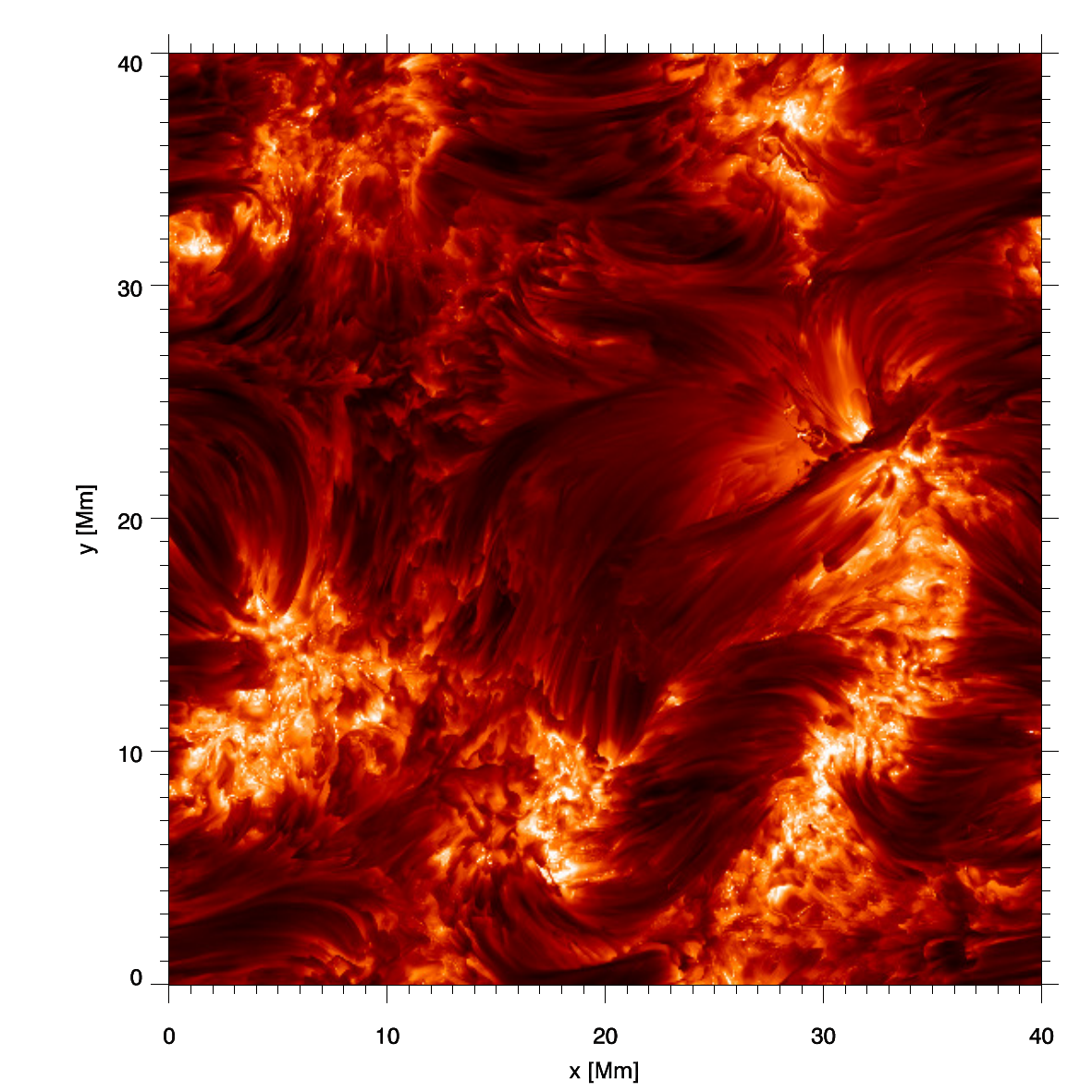}
\caption{Brightness temperature in the nominal line core of H$\alpha$ in a simulated active region. The radiation-MHD simulation was performed with the \textsc{MURaM} code by Danilovic et al. (2019). The simulation used a single-bin LTE treatment of chromospheric radiative transfer. Nevertheless, the simulation resembles observations rather well. The 3D non-LTE radiative transfer calculation used to obtain the H$\alpha$ intensity was done with the Multi3d code.}
\label{fig:sanja_plage_halpha}     
\end{figure}

Radiative transfer and non-equilibrium ionisation in the chromosphere is probably the least studied and the methods described in Sections~\ref{sec:chromosphere} and~\ref{sec:noneq-ionisation} are likely the most inaccurate of all method used in the photosphere-chromosphere-corona system. {A} long sequence of approximations {is}  needed to order to make the problem computationally feasible. Testing the accuracy of all approximations combined can only be done in 1D geometry (using RADYN or similar codes), and has so far not been done. A particular worry is the crude way in which the radiative transfer in the Ly-$\alpha$ line is implemented in the non-equilibrium ionisation method,

Interestingly, the way how chromospheric radiative transfer is implemented, and the accuracy of the method used seems however to have only a minor effect on the overall structure of the simulated chromospheres. A \textsc{MURaM} simulation of active regions that only contains single-bin (gray) LTE radiative transfer and a coronal loss function produce chromospheres whose structure resembles the real chromosphere (see for example 
\cite{2019arXiv190601098B}
and Fig.~\ref{fig:sanja_plage_halpha}). 

I argue that this is a consequence of the physics of the chromosphere in the MHD approximation. Thermal conduction in the chromosphere is not efficient, and the only way a Lagrangian fluid element can lose energy is through radiation. In a chromosphere that is in a statistically stable state, the dissipation of non-thermal energy into heating and the radiative cooling must be in balance in a volume and time-integrated sense. An increase in non-thermal energy deposition must be accompanied by an increase in radiative losses. In other words: radiative losses are a reaction to heating, and the chromosphere will adapt its thermodynamic state until heating and cooling are in balance again. 

Here is the catch: the radiative losses are very sensitive to temperature, but the mechanisms that dissipate non-thermal energy are not. The radiative losses assuming LTE scale as $T^4$, while the radiative losses in the coronal equilibrium approximation scale exponentially with temperature under chromospheric conditions (see Fig.~\ref{fig:radiative-loss-function}). The non-LTE chromospheric loss function will lie between these two extremes most of the time. 

In the MHD approximation there are only two mechanisms that {irreversibly} convert non-thermal energy into heat: viscosity and electrical resistance. The first one scales as $T^{1/2}$ (assuming a dilute gas of rigid elastic spheres), while the second scales as $\ln T/T^{-3/2}$ (assuming Spitzer resistivity). In practice the viscosity and resistance are replaced by numerical terms that are independent of temperature in almost all numerical simulations.

When the description of the chromospheric radiative losses or the equation of state is not correct, then the simulation will change the temperature compared to the correct solution until balance between energy gains and losses is achieved again. The steep temperature dependence of the radiative losses makes the temperature change modest (say 1000~K--2000~K if our methods are not too bad), but the effect on the non-thermal energy dissipation rate is small. The result is a chromosphere with the wrong temperature, but almost correct dissipation, density and velocity structure.

It thus seems that one can study certain aspects of chromospheric physics without a too complicated treatment of the radiative losses and equation of state\footnote{Simulations including ambipolar diffusion are an exception \citep{2012ApJ...753..161M,2018A&A...618A..87K}. {The ambipolar diffusivity depends explicitly on the number density of neutrals and ions, and a correct non-equilibrium treatment of the ionisation is essential to compute the diffusivity.}}. 
This does however not mean that the work described in Sections~\ref{sec:chromosphere} and~\ref{sec:noneq-ionisation} can be ignored. The only proper way to validate models is to compute the various diagnostics (line profiles and continua) and compare to observations. The emergent intensities in the diagnostics sensitively depend on the temperature and electron density. Comparison of models with observations 
\citep[such as in][]{2013ApJ...772...90L,2018A&A...611A..62B}
show that current models are not getting the intensities right, and that they need to be refined. The treatment of chromospheric radiative energy exchange is very likely one of the aspects in need of further improvement.

The reverse comparison is also true: inferred model atmospheres generated through non-LTE inversions of observations
\citep[e.g.,][]{2017SSRv..210..109D,2019A&A...623A..74D}
can only be used to constrain physical models of the chromosphere if we are sure that the treatment of radiation in the models is done sufficiently accurately. 

So what can be improved? I propose the following non-exhaustive list:
\begin{itemize}
\item It would be interesting to test, and if necessary improve, the accuracy of Skartlien's multi-bin radiative transfer with scattering in the mid and upper chromosphere. If successful, this would increase the accuracy compared to the radiative loss tables of \citet{2012A&A...539A..39C}, which ignore 3D effects.
\item The tables for \CaII\ and \MgII\ of \citet{2012A&A...539A..39C} should be updated. They were calibrated on a single 2D radiation-MHD simulation with a weak magnetic field using 1.5D radiative transfer. A much larger variety of models is available now, and 3D non-LTE radiative including PRD is now possible \citep{2017A&A...597A..46S}. 
\item The accuracy of the non-equilibrium ionisation methods should be more critically assessed. Comparison against the full physics is only possible in 1D but this will already give more insights in the possible deficiencies of the model, and in particular of the treatment of Ly$\alpha$.
\item The escape probability method for approximating the chromospheric radiation field discussed in Sect.~\ref{sec:escape-prop} should be tested and developed further.
\item It should be investigated whether non-equilibrium ionisation of elements that are important for radiative losses in the TR and corona can be implemented in a sufficiently computationally efficient fashion. The ionisation balance of helium can be modeled with only one energy level per ionisation stage and parametrized ionisation/recombination rates that include the effects of the excited levels. If a similar thing can be done for other elements, then inclusion of non-equilibrium effects is otherwise straightforward.
\end{itemize}
Let us hope that improvements along these lines, as well as others, will be presented during the coming years.

\begin{acknowledgements}
This work was supported by grants from the Knut and Alice Wallenberg foundation (2016.0019) and the Swedish Research Council (2017-04099) and benefitted from  discussions within the activities of team 399 `Studying magnetic-field-regulated heating in the solar chromosphere' at the International Space Science Institute (ISSI) in Switzerland.
\end{acknowledgements}

\bibliographystyle{spbasic}   

\bibliography{radhydro}   

\end{document}